%% file: main.tex
\documentclass[a4paper,fleqn]{cas-sc}

\usepackage[authoryear,longnamesfirst]{natbib}
\usepackage{xcolor}
\usepackage{subcaption}
\usepackage{graphicx}
\usepackage{amsmath}

\newcommand{\cs}[1]{\textcolor{blue}{[\textit{#1} --CS]}}

\begin{document}

\let\WriteBookmarks\relax
\def\floatpagepagefraction{1}
\def\textpagefraction{.001}
\renewcommand\thefigure{\thesection.\arabic{figure}}  

\shorttitle{Modeling road user response timing in naturalistic settings: a surprise-based framework}    

\shortauthors{\textcolor{white}{-}}  

\title [mode = title]{Modeling Road User Response Timing in Naturalistic Traffic Conflicts: A surprise-based framework}

\author[1]{Engstr{\"o}m, Johan}
\author[1]{Shu-Yuan Liu}
\author[1]{Azadeh Dinparastdjadid}
\author[1]{Camelia Simoiu}



\affiliation[1]{organization={Waymo LLC},
            addressline={1600 Amphitheatre Parkway}, 
            city={Mountain View},
            postcode={94043}, 
            state={CA},
            country={USA}}

\begin{abstract}
There is currently no established method for evaluating human response timing across a range of naturalistic traffic conflict types. 
Traditional notions derived from controlled experiments, such as perception-response time, fail to account for the situation-dependency of human responses and offer no clear way to define the stimulus in many common traffic conflict scenarios.
As a result, they are not well suited for application in naturalistic settings. 
We present a novel framework for measuring and modeling response times in naturalistic traffic conflicts applicable to automated driving systems as well as other traffic safety domains. 
The framework suggests that response timing must be understood relative to the subject's current (prior) belief and is always embedded in, and dependent on, the dynamically evolving situation. 
The response process is modeled as a belief update process driven by perceived violations to this prior belief, that is, by surprising stimuli. 
The framework resolves two key limitations with traditional notions of response time when applied in naturalistic scenarios: 
(1) The strong situational-dependence of response timing and 
(2) how to unambiguously define the stimulus. Resolving these issues is a challenge that must be addressed by any response timing model intended to be applied in naturalistic traffic conflicts. 
We show how the framework can be implemented by means of a relatively simple heuristic model fit to naturalistic human response data from real crashes and near crashes from the SHRP2 dataset and discuss how it is, in principle, generalizable to any traffic conflict scenario. 
We also discuss how the response timing framework can be implemented computationally based on evidence accumulation enhanced by machine learning-based generative models and the information-theoretic concept of surprise.  
\end{abstract}

\begin{keywords}
response time  \sep  
human behavior modeling \sep 
naturalistic driving data \sep  
autonomous vehicles \sep
surprise
\end{keywords}

\maketitle

\section{Introduction}\label{intro}
\input{1_intro}

\section{Measuring and modeling response timing in the wild}
\input{2_measuring_respt}

\section{Methods}
\input{3_methods}

\section{Results}
\input{4_results}

\section{A computational implementation}\label{sec5_computational}
\input{5_computational}

\section{Discussion and conclusions}
\input{6_discussion}

\bibliographystyle{cas-model2-names}


\bibliography{references}



\appendix
\balance
\input{appendix_A}
\newpage
\input{appendix_B}
\input{appendix_C}

\end{document}

%% file: 1_intro.tex
\noindent
The analysis and modeling of response timing has a long history in psychology, cognitive science, and neuroscience (e.g.,~\cite{medina2015advances}, \cite{posner1978chronometric,welford1952psychological}). It is also an important topic in applied traffic safety research, where it has been used extensively in road design (\cite{fambro1998driver}), forensics (\cite{maddox2012looming}), driver distraction research (\cite{angell2006driver,engstrom2017effects}), and advanced driving assistance and driving automation system evaluation (\cite{gold2013take,lee2002collision,aust2013effects,mcdonald2019toward}).

In the context of automated driving systems (ADS), response timing models may be used to benchmark ADS response performance in traffic conflicts and model how other agents respond to the ADS in simulation-based evaluations.
The performance of an attentive and non-impaired driver in a crash-imminent situation presents a natural reference point on the human driving spectrum to help assess autonomous vehicle performance, although this does not exclude other types of collision avoidance benchmarks not directly based on human performance. 
Such \textit{behavior reference models} can be used to define requirements on maximum ADS response latencies (i.e., the time it takes for an ADS to compute and initiate a response to a sensor input) and provide performance benchmarks for ADS collision avoidance evaluation.\footnote{See the UN regulation on Automated Lane Keeping Systems, UN, 2021, for an existing example of an ADS collision avoidance reference model based on human performance.} 
More specifically, the present paper discusses a behavior reference model for response timing that represents a Non-Impaired road user with their Eyes ON the conflict (NIEON).
The paper thus focuses exclusively on the response timing component of collision avoidance (i.e., the process of deciding whether to brake and/or swerve in response to a traffic conflict) and does not address the modeling of the evasive maneuver (i.e., braking and/or swerving profiles), which is in itself a complex and challenging topic. While the present paper focuses mainly on ADS benchmarking applications, the proposed framework is applicable to any type of road user response timing modeling, as further addressed in the Discussion section. 

Since ADSs operate in a variety of real-world situations, a response timing model for ADS needs to be applicable across a wide range of naturalistic traffic conflicts, where \textit{traffic conflict} here refers to ``a situation where two or more conflict partners approach each other in time and space to such extent that a crash is imminent if their movements remain unchanged''\footnote{
This corresponds to the notion of a trajectory conflict in ISO/TR 21974-1:2018(E), 
which suggests a broader definition of traffic conflicts also including run-off road events and proximity conflicts.} (\cite{svensson2006estimating}).
Traditional approaches for analyzing response times have typically been developed in the context of controlled experiments (\cite{posner1978chronometric,olson1989driver}), with clearly defined stimuli and specific task instructions. 
However, as further discussed below, due to the often ambiguous and continuous nature of the stimuli that road users respond to in real traffic conflicts, and the strong situation dependence of response timing, these traditional models are difficult to apply in naturalistic scenarios. 
While more recent evidence accumulation models (e.g., \cite{markkula2014modeling,markkula2016farewell,engstrom2018simulating,dinparastdjadid2020modeling}) resolve many of these issues, they have so far only been applied to specific scenarios and stimulus types (in particular rear-end scenarios and visual looming).
As far as we are aware, there currently exists no comprehensive framework for analyzing and modeling human response performance across a broad range of conflict scenarios in the wild. 

The key contribution of this paper is the proposal of such a framework. 
The proposed framework casts evidence accumulation in a \textit{predictive processing} framework (\cite{clark2013whatever,clark2015surfing,clark2023experience,engstrom2018great}). 
More specifically, we propose to model response timing in road traffic based on \textit{belief updating driven by surprising evidence}, and demonstrate a practically feasible means to implement it.

The next section provides a review of traditional notions of response time used in road user behavior and traffic safety research and their key limitations when applied in real-world naturalistic driving scenarios. 
To address these limitations, we present our proposed response modeling framework and demonstrate how it can be implemented in practice with a relatively simple model based on human annotation guided by heuristic rules, fit to human response data. 
Section 3 describes the methods used to select naturalistic rear end crashes and near crashes for the analysis and extract the signals needed for the analysis, and Section 4 presents the modeling results. 
Section 5 presents some general ideas, and a specific simulation example, on how the framework may be implemented computationally. 
The paper concludes with a general discussion, conclusion, and directions for further research.

%% file: 2_measuring_respt.tex
In this section, we first briefly review traditional notions of response time and discuss why they are difficult to apply in naturalistic scenarios.
We then introduce the proposed framework and outline a practically feasible way to implement it based on manual annotation of the stimulus guided by scenario-specific heuristic annotation rules.

\subsection{Traditional notions of perception-response time and their limitations}

Traditionally, applied work on human response times in traffic conflicts typically starts from the notion of a response process divided into a sequence of information processing stages, such as detection, identification, decision and response (e.g., \cite{olson1989driver}).
The time it takes for the brain to execute this process is typically referred to as perception-response time (PRT).
This type of model has its basis in a long and rich research tradition in cognitive psychology known as mental chronometry, which refers to the scientific study of reaction time as the basis for inferring cognitive mechanisms (e.g.,~\cite{medina2015advances,posner1978chronometric,welford1952psychological}). 

\begin{figure}[t]
	\centering
		\includegraphics[width=0.5\textwidth]{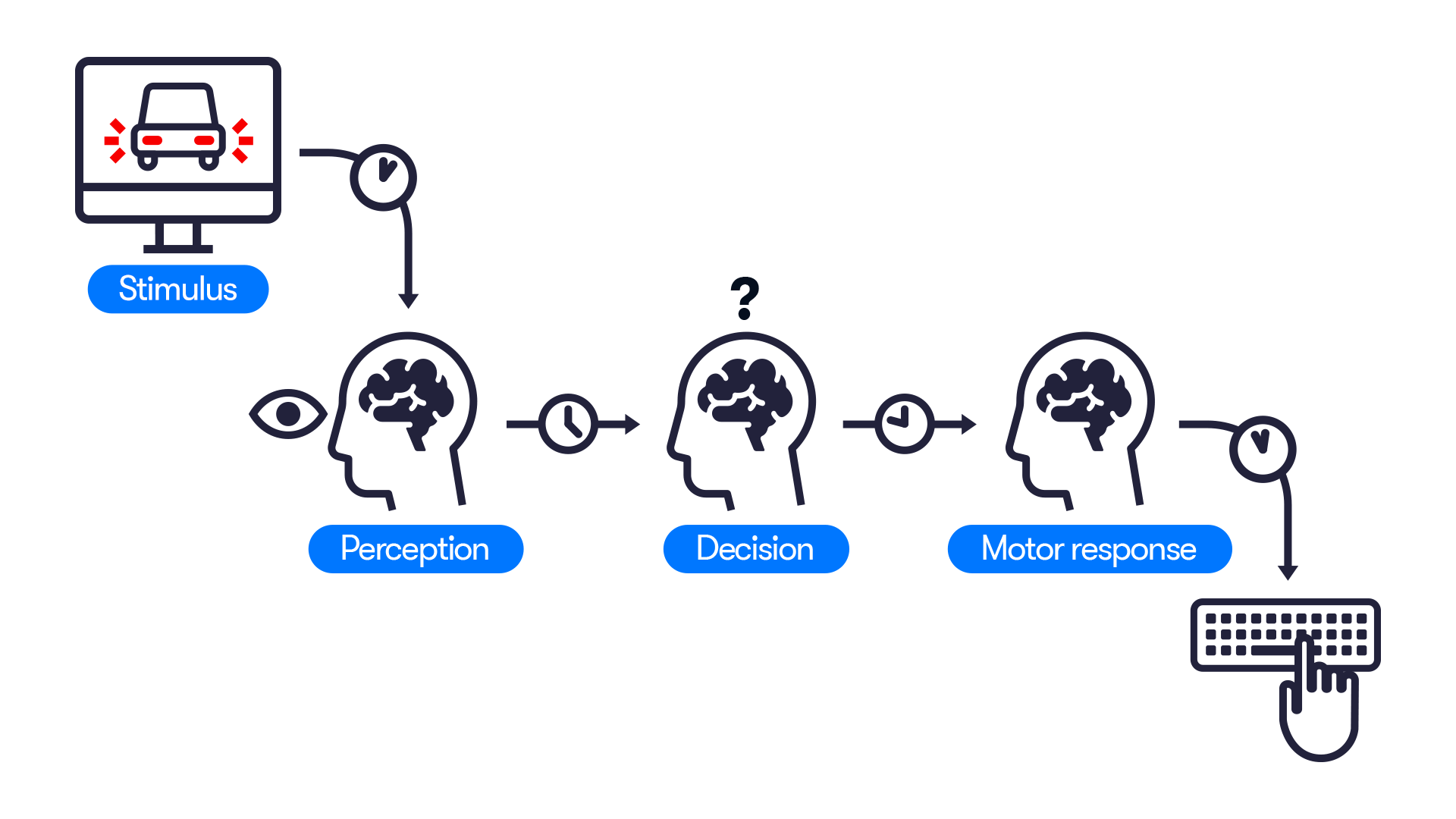}
	  \caption{The traditional notion of response time as determined by the time to process a stimulus through a set of sequential stages in the brain.}\label{brain_processing}
\end{figure}

PRT can be generally conceptualized as the time it takes to process a well-defined stimulus through a sequence of processing stages in the brain (see Figure~\ref{brain_processing}).
PRT is thus typically seen as a property ascribed to the individual agent, as opposed to being a property of the dynamically evolving situation as a whole.
Based on this tradition, ~\cite{green2000long}, in an influential paper, proposed a set of canonical driver response times based on the driver's degree of expectancy of the event: ``expected'' events (0.7-0.75s response time), ``unexpected'' events (1.25s) and ``surprise'' intrusions (1.5s). 

This type of model is problematic to apply in naturalistic traffic scenarios, where road users behave without instructions, for two main reasons:  (1) The strong situation-dependence of the response process, and (2) lack of a principled, generalizable means to define the stimulus.

\textbf{Situatedness of the response process}.
As discussed in ~\cite{markkula2016farewell}, response timing in traffic conflicts depends strongly on the dynamically evolving situation. 
For example, as pointed out by~\cite{summala2000brake}, criticizing Green's attempt to define canonical driver-centric response times mentioned above, response time correlates strongly with the urgency of the scenario. 
A strong relationship between scenario kinematics and response timing has since been demonstrated in numerous studies, both in naturalistic driving data (\cite{markkula2016farewell} and in controlled experiments (e.g.,~\cite{aust2013effects,li2014effectiveness,bianchi2020drivers}, and a meta-analysis of automation take-over studies in~\cite{mcdonald2019toward}). 
A meta-analysis of published experimental (simulator and field) studies measuring response times in lead vehicle braking events showed that the observed average response times varied widely between studies (0.5s to 1.5s) and that most (88\%) of this variance could be explained by the scenario kinematics implemented in the different studies (specifically the time gap between the following and the lead vehicle at the moment the lead vehicle started braking (\cite{engstrom2010scenario,engstrom2018simulating}).
A model based on the traditional (agent-centric, situation-independent) PRT concept, such as~\cite{green2000long}'s model, would predict a constant response time across scenarios with different kinematics (e.g., rear end conflict scenarios with different initial time gaps) and thus fails to account for this variance.
Hence, naturalistic road user responses to traffic scenarios are to a large extent determined by the dynamically evolving situation and cannot be modeled solely in terms of information processing inside the head of the road user.

\textbf{Defining the stimulus}.
As outlined above, PRT relies on a clear-cut definition of the stimulus and its onset (i.e., when to start the clock in Figure~\ref{brain_processing}).
In the laboratory, the stimulus and its onset point can be conveniently defined by the experimenter and implemented as part of the instructions to participants (e.g., ``press the brake pedal when you see a brake light onset''). 
However, in naturalistic settings, it is often unclear how to define the stimulus that a road user is supposed to respond to (e.g.,~\cite{green2009perception, fajen2006learning}). 
Whether a road user will respond to a stimulus at all depends critically on their current expectation of how the situation will play out. 
For example, when a lead vehicle and a following vehicle are approaching a red light, the brake light onset of the lead vehicle is typically expected by the following driver (who may have already started slowing down prior to the brake light onset). 
In such a situation it makes little sense to treat the brake light onset as the start of a perception response time.
This is similar to the ``tollbooth problem'' introduced by~\cite{fajen2006learning}: when a driver approaches a tollbooth, it makes little sense to assign a stimulus onset for perception response time to the point where the toll both is first seen by the driver, as there is typically little reason for the driver to hit the brakes at this point.

\begin{figure}[t]
	\centering
		\includegraphics[width=0.5\textwidth]{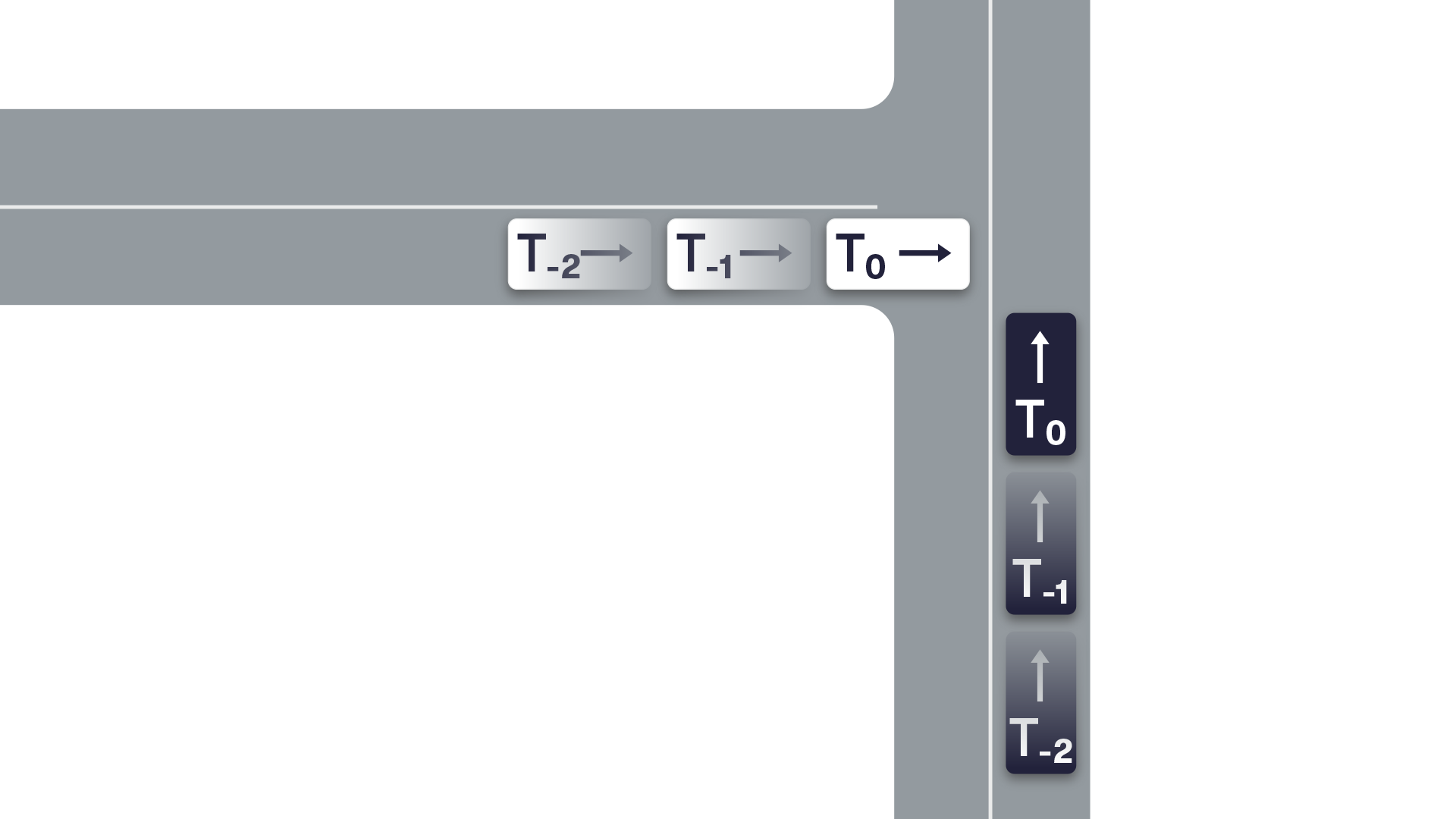}
	  \caption{Example of a gradually evolving scenario without a physically well-defined stimulus onset. The white car is referred to as the Principal Other Vehicle (POV) in the text and the black car as the Subject Vehicle (SV).}\label{gradual_scenario}
\end{figure}

Moreover, even if a stimulus signal can be unambiguously identified, well-defined stimulus onsets are rather the exception than the rule in naturalistic traffic conflicts. 
Traffic situations often evolve gradually which makes it difficult to determine exactly when to start the clock for measuring a response time. 
In addition, naturalistic situations often lack a distinct physical feature that can be used to define a stimulus onset.
For example, consider the scenario depicted in Figure~\ref{gradual_scenario}. 
Here, the subject  vehicle (SV, black car) is driving on a main road with the right of way and another vehicle (the Principal Other Vehicle, POV; white car) approaches on a perpendicular intersecting road at constant speed. 
Instead of yielding, as would be expected, the POV continues ahead and enters in front of the SV. 
The POV  was visible to the SV driver long before the scenario turned critical and, when analyzing the response performance of the SV driver, it is not immediately clear at what point during the POV's  approach to the intersection to set the stimulus onset.
It should be noted that this problem is not unique to naturalistic settings; if the scenario in Figure~\ref{gradual_scenario} would be implemented in a driving simulator, it would be equally hard to unambiguously define the stimulus onset.
However, in driving simulator studies (and real-world controlled experiments) the experimenter can design the scenarios to avoid this problem, a luxury that is not afforded by naturalistic settings.

An additional concern is when to stop the clock, as the driver's braking or steering response may also occur gradually and, as demonstrated by~\cite{markkula2016farewell}, the magnitude of the response also depends on the situation kinematics. 
However, this issue is not further addressed here as the scope of the paper is limited to response timing, not the type of evasive maneuver or the details of how it is performed. 

To summarize, the traditional notion of perception-response time is based on concepts mainly applicable in controlled experiments and hard to unambiguously apply in naturalistic settings where the stimulus onset is not well-defined and road users do not behave based on instructions.
In particular, traditional approaches do not account for the situatedness of the response process and suffer from the lack of a principled way to define the stimulus and its onset in naturalistic situations. 
Any model that claims to represent human response timing in naturalistic traffic scenarios must articulate how these two issues are being addressed. 
In what follows, we propose a solution based on existing evidence accumulation models and the notion of predictive processing, casting road user response timing in terms of a belief updating process driven by surprising stimuli.

\subsection{A framework for modeling road user response timing in the wild}

This section describes our proposed conceptual framework for response time modeling.
The key idea is that a road user's response to a traffic event can be conceptualized as a process of inference, where the agent's behavior is initially guided by prior beliefs about the causes of sensory input, which are continuously updated to posterior beliefs based on the accumulation of new evidence.
Note that, while we use concepts from Bayesian statistics to explain the framework, implementations of the inference and belief updating processes do not necessarily have to be Bayesian. 

\begin{figure}[t]
    \centering
       \begin{subfigure}[b]{0.49\textwidth}
       \includegraphics[width=0.95\textwidth]{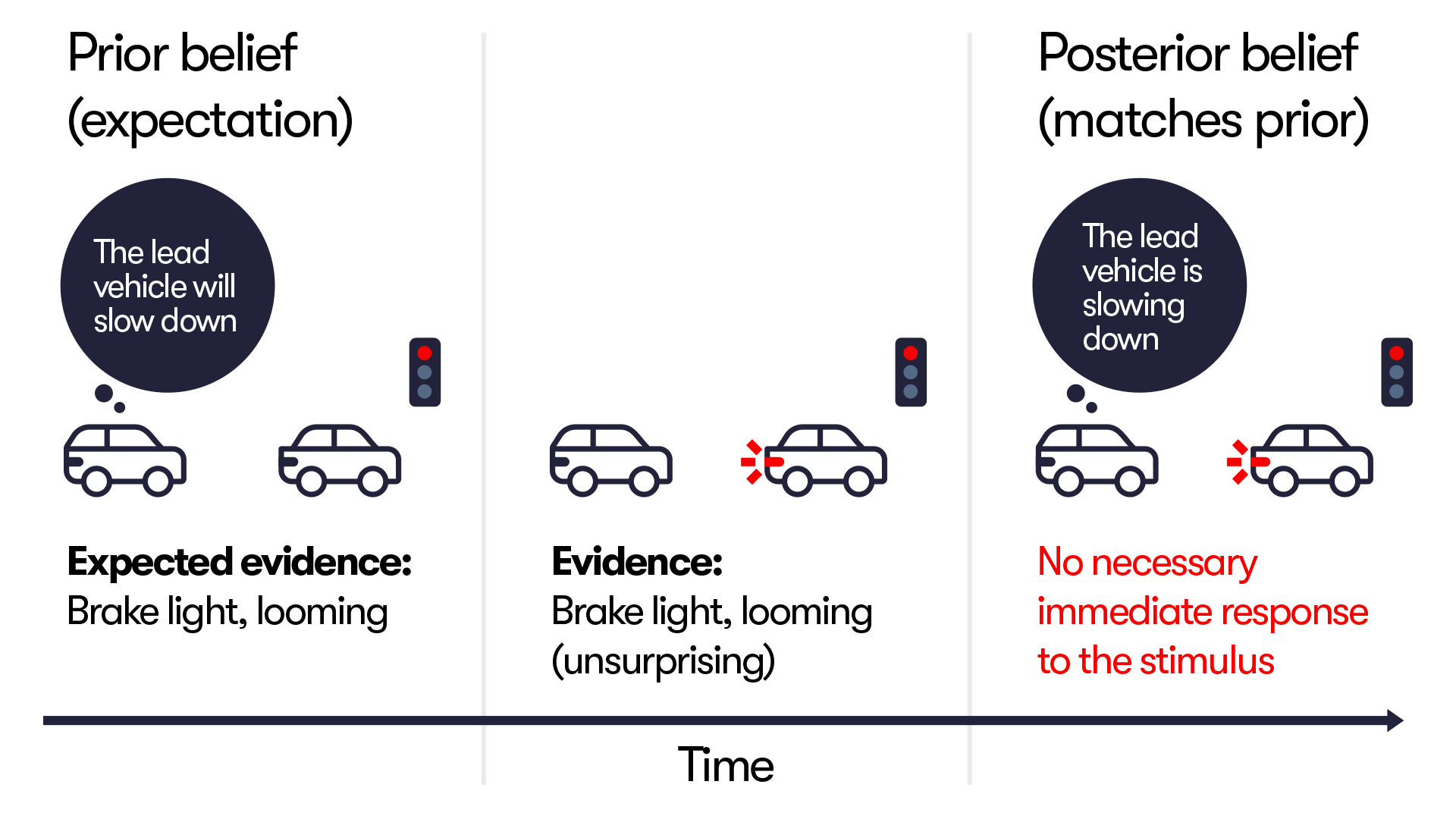}
    \end{subfigure}
    
    \begin{subfigure}[b]{0.49\textwidth}
       \includegraphics[width=0.95\textwidth]{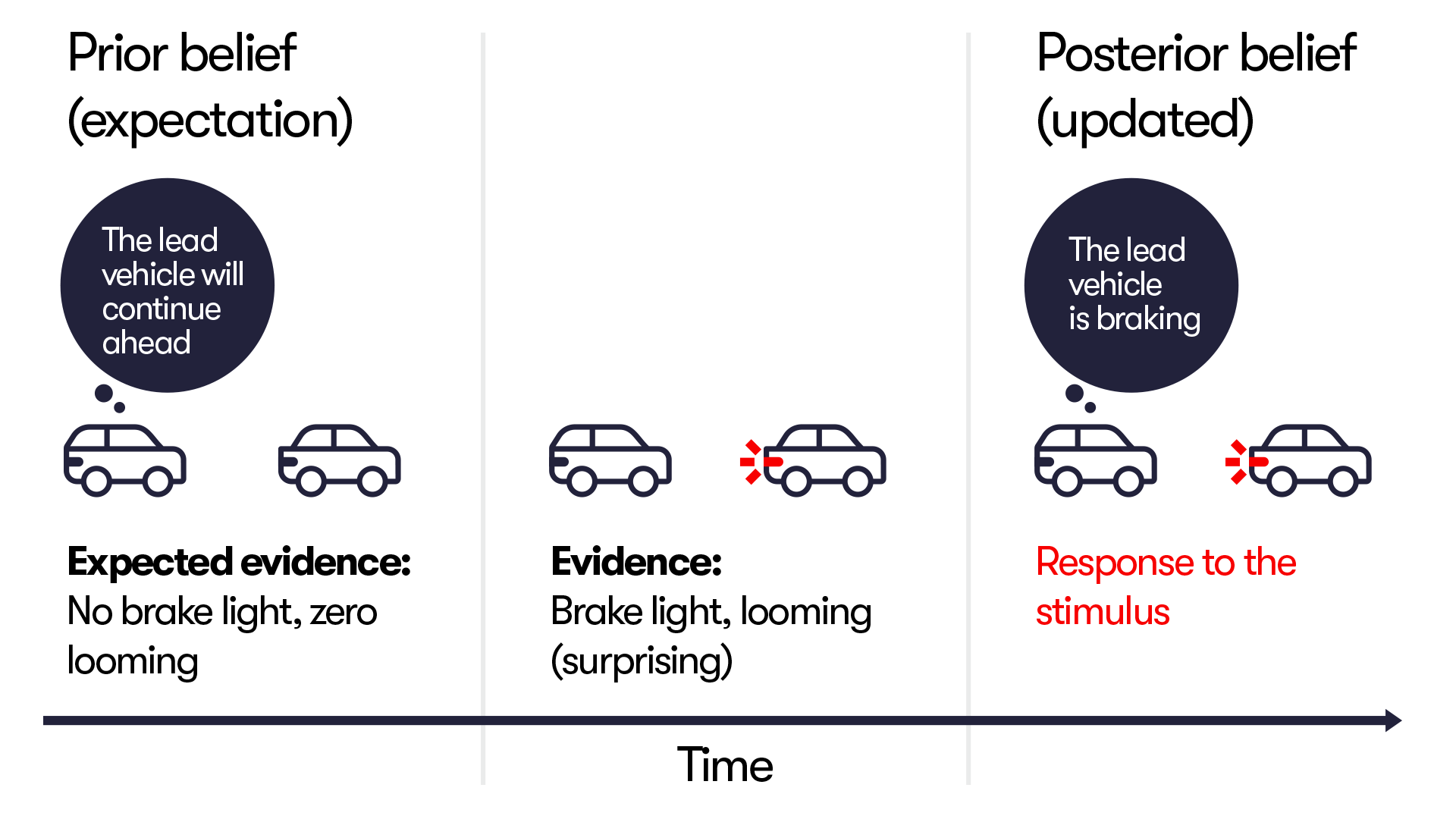}
    \end{subfigure}
\caption{Conceptual illustration of a belief update process.
Top: The evidence is consistent with the prior belief and, hence, the posterior belief remains the same.
Bottom: The evidence does not match the prior belief and is thus surprising, leading to an update of the posterior belief. In a traffic conflict, where imminent action is needed to avoid collision, the belief update will lead to an evasive maneuver (e.g., braking and/or steering). In the current framework, response timing is understood in terms of this belief updating process.}\label{belief_updating_process}
\end{figure}

This belief updating process is schematically illustrated in Figure~\ref{belief_updating_process}, where a driver is following a lead vehicle and receives evidence about whether or not the distance to the lead vehicle is closing.
The evidence here consists of brake light onsets and visual looming (the optical expansion of the lead vehicle, typically operationalized as the rate of change of the angle, $\theta$, that an object subtends on the subject's retina, $\theta$-dot,~\cite{lee1976theory}). 
In the top panel, the driver's prior belief is that the lead vehicle will slow down, in which case the expected evidence (brake lights on and looming) matches the actual evidence. 
Thus, the belief does not change and the posterior matches the prior. 
This represents an unsurprising situation in which the following driver continues to act upon their (correct) expectations and slow down as needed. 
The following driver's slowing down behavior is then largely proactive and does not necessarily involve any response to the behavior of the lead vehicle at all (i.e., only a planned deceleration based on the prior expectation that the lead vehicle will slow down based on the red light and the traffic situation as a whole). Alternatively, the following driver may wait to slow down until the lead vehicle starts braking. 
In this case, the behavior is triggered by an anticipated cue (e.g., the brake light onset of the lead vehicle). 
However, this still represents a largely unsurprising situation playing out as initially expected which does not require a radical change in belief from the following driver. 

\begin{figure}[t]
	\centering
		\includegraphics[width=0.5\textwidth]{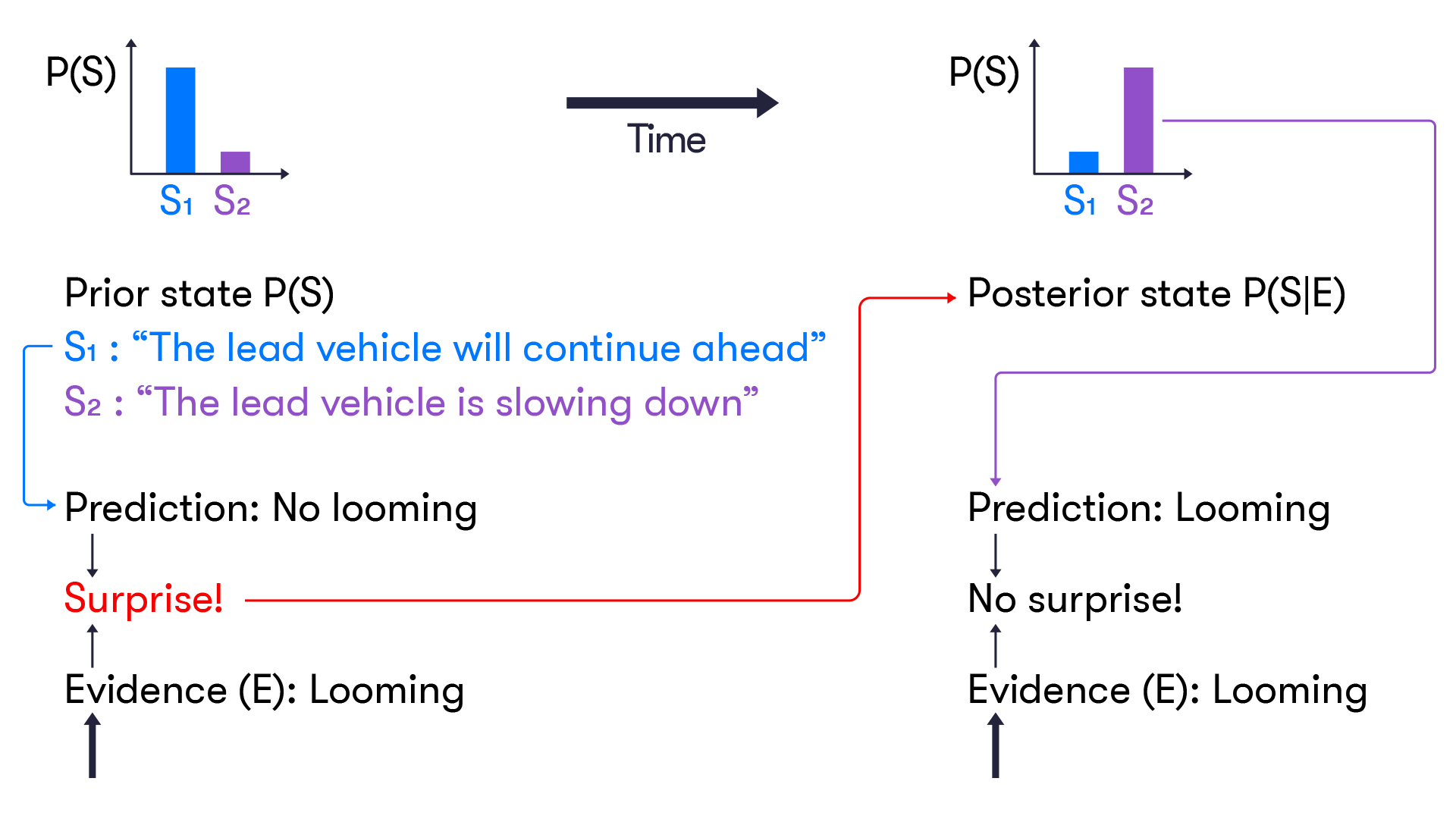}
	  \caption{Formal illustration of the proposed belief update process for the surprising case in Figure~\ref{belief_updating_process}}
	  \label{formal_belief_updating}
\end{figure}
  
However, if the evidence is inconsistent with the prior belief, the evidence is surprising and the posterior no longer matches the prior. 
This situation is illustrated in Figure~\ref{belief_updating_process} (bottom panel), where the driver initially expects the lead vehicle to continue ahead but the lead vehicle surprisingly brakes, as evidenced by brake lights and looming. 
The posterior belief then needs to be updated to match the new evidence. 
If the situation is not urgent (e.g., the braking lead vehicle is far ahead) immediate action may not be needed and just updating the belief may be sufficient given that the driver can implement a new proactive plan of action at a later stage.
However, if the situation is urgent, that is, a traffic conflict, the belief update will generate an immediate action, in this case braking and/or swerving to avoid a collision.
Hence, in naturalistic scenarios, where no instructions are given to drivers to respond to certain stimuli, the concept of response time is only meaningfully defined for traffic conflicts, since in non-conflict scenarios there is no need to respond immediately in the first place (cf. the toll-both problem referred to above).
It should be noted that the term ``belief'' is here used in an abstract sense (technically, a probability distribution over possible outcomes as described below) and does not necessarily have to involve conscious thought (as may seem to be implied by the conceptual illustration in Figure~\ref{belief_updating_process}).

The idea that perception can be understood as inference dates back to~\cite{helmholtz1876origin} and the general notion that cognition and behavior can be understood in terms of Bayesian inference and prediction error minimization is today an influential paradigm in cognitive neuroscience, under various names such as the \textit{Bayesian brain} (\cite{knill2004bayesian}), \textit{predictive processing} (\cite{clark2013whatever, clark2015surfing, clark2023experience}) and \textit{active inference} (\cite{friston2017active,parr2022active}). This framework has also recently been applied in the context of road user behavior modeling (\cite{engstrom2018great,kujala2021inattention,wei2022modeling,wei2022worldmodel, wei2023carfollowing})

A more formal account of the proposed belief updating scheme is illustrated in Figure~\ref{formal_belief_updating}, continuing the example from Figure~\ref{belief_updating_process} (bottom). 
Here, the following vehicle driver’s prior belief about whether the lead vehicle will continue ahead or brake is represented as probability distribution P(S), where S represents the state of the lead vehicle. 
Initially, the belief that the lead vehicle will continue ahead ($S_1$) dominates the prior distribution, which entails the prediction that the observed looming evidence (E) will remain close to zero. 
When the lead vehicle slows down, the observed looming deviates from the predicted looming and generates a prediction error or surprise. As described in (\cite {dinparastdjadid2023surprise}), surprise can be quantified in different ways, for example in terms of surprisal or Shannon surprise (the negative log probability of the observation under the predicted distribution, see Section~\ref{5_computational}) or Bayesian surprise (the Kullback-Leibler divergence between the prior and posterior belief distribution~\cite{itti2009bayesian}, which drives an update of the prior belief.

Based on this framing, road user response timing in a traffic conflict can be understood in terms of the time course of the belief updating process (i.e., the variation in beliefs over time).
This is conceptually illustrated in Figure~\ref{heuristic_labels_belief}, where the evasive maneuver is generated once the posterior belief reaches a decision threshold. 
As discussed in~\cite{bogacz2006physics,dinparastdjadid2020modeling}, the decision threshold represents a trade off between speed and accuracy.
If the threshold is set too low, the agent will respond fast but may overreact in situations that may not require an evasive maneuver. 
Conversely, if the threshold is set too high, the agent may respond too late.
In terms of the current framework, the purpose of the evasive maneuver is to eliminate the surprise and bring the agent back to its preferred, non-conflict, state. Such surprise-minimizing evasive action can, in line with the present framework, be conceptualized and modeled in terms of active inference (\cite{engstrom2018great,friston2017active,wei2022modeling,wei2023carfollowing,parr2022active}).
However, since the focus of the present paper is specifically on response timing, this topic is not further addressed here.  

As noted above, and reviewed by~\cite{olson1989driver, green2000long}, road users' response performance is influenced by a range of endogenous and exogenous factors such as individual characteristics (age, driving experience), temporal agent states (e.g., visual distraction, cognitive load, fatigue alertness) and stimulus conspicuity (e.g. affected by darkness, sunlight glare, fog and rain). 
In the current framework, the effect of such factors can be modeled as influencing (speeding up or slowing down) the belief updating process in different ways (see~\cite{engstrom2018great,markkula2017simulating} for response timing models of the effects of cognitive load high/low arousal based on this general idea).  

To summarize, in the proposed framework, responses to traffic conflicts are driven by surprising evidence (i.e., observations that are not explained or predicted by the prior beliefs). 
This has the important implication that response timing is always relative to the prior belief of the agent. 
The prior belief is both context-dependent (i.e., dependent on the specific scenario) and determined by the road user's prior experience. 
For example, US and French drivers would likely have somewhat different prior beliefs in a given scenario, given that they have been exposed to different driving environments and traffic cultures, which may affect how they respond to a conflict. 
The heuristic model described in the next section does not account for such individual factors, but they can, at least in theory, be captured by machine-learned generative models trained on data from different driving environments (e.g., San Francisco vs. Paris, see Section~\ref{sec5_computational} and (\cite {dinparastdjadid2023surprise}) ). 

Another key idea behind the proposed framework is that response timing can be understood in terms of the time course of a belief updating process. 
This belief updating can be represented computationally as a process of evidence accumulation where the strength of the belief grows towards a decision/response threshold over time as surprising evidence is gathered over time. 
response timing thus depends on the rate of the incoming surprising evidence. 
If the evidence comes in at a fast rate (e.g., the lead vehicle brakes hard), the surprise signal will grow fast, driving a fast belief update and a quick response. Conversely, a more gradually developing scenario will lead to a longer response time. 

This type of model, in line with related existing evidence accumulation response models (e.g,~\cite{markkula2014modeling,markkula2016farewell,engstrom2018great,dinparastdjadid2020modeling}), thus offers a natural explanation for the strong situatedness of response timing reported in the literature.
Additionally, the framework offers a natural way to define the stimulus in naturalistic scenarios in terms of surprising evidence, and the stimulus onset as the first surprising observation.
This notion accounts for situations such as that illustrated in Figure~\ref{gradual_scenario}, which lack a well-defined physical stimulus onset marker and the onset is solely defined by the (surprising) violation to a prior belief. 
In the scenario in Figure~\ref{gradual_scenario}, the prior belief  is that the approaching vehicle will slow down and yield and the accumulating surprising evidence is the observation that the vehicle continues ahead at constant speed so that it can no longer brake comfortably to a stop.
From one perspective, the proposed framework can be seen as eliminating the concept of response time altogether, focusing instead only on modeling the response \textit{onset}, as in existing evidence accumulation models referred to above (\cite{markkula2016farewell}, and our computational model described in Section~\ref{sec5_computational}, Figure~\ref{fig_computational}). 
However, the framework also allows for retaining a notion of (situation-dependent) response time which may be useful in certain applications, such as the heuristic model described in the next section. 

It should be noted that both the heuristic and computational model implementations presented in this paper only apply to traffic conflicts, that is, urgent situations where an imminent response is required.
This is because the model only represents a single belief update leading to a single-step emergency evasive maneuver (e.g., slamming on the brakes). 
However, the surprise-based evidence accumulation framework in general allows for more advanced models that respond in a more gradual or step-wise fashion to a surprising event (see~\cite{svard2017quantitative,svard2021computational} for an example of such a model operating on prediction error).

In addition to accounting for the key limitations of the traditional perception-response time concept, a key advantage of the proposed framework is its generality. Models based on the present framework can, in principle, be applied to any type of traffic conflict given that the prior belief, the posterior belief and the surprising evidence that drives the belief update can be defined. 
The parameters of  response timing models based on these general principles can then be fit to human response data to represent the  response performance of the human cohort that the data is sampled from (e.g., non-impaired drivers with their eyes on the conflict). 
Ideally, a fully computational implementation of the belief updating process would be desired, and we present an example of such a model implementation in Section~\ref{sec5_computational}, where the prior belief is automatically generated by a machine-learned generative model. 
However, there are still a number of practical challenges associated with implementing these types of generalized computational response models. 
In particular, publicly available naturalistic traffic conflict data such as SHRP2 (see below) is typically limited in terms of the environment sensor data available, for example the precise location of other road users, which precludes the application of of machine-learned generative models.
Other sources of existing naturalistic conflict data may be of even lower fidelity than SHRP2, only including dash cam footage and few basic kinematic sensors. 
Thanks to its generality, the present framework can be implemented in simpler ways, making use of human annotations based heuristic guidelines for annotating the prior and posterior beliefs, and the surprising evidence, in a given scenario. 
The following sections describe how such a simplified heuristic model can serve as a proxy for a full-blown computational model, thus making it feasible to evaluate response timing across a range of naturalistic scenarios with data of varying fidelity.

\subsection{Implementing the framework in practice: A heuristic model for ADS response time benchmarking}

We describe a practically feasible approach for implementing the response modeling framework described above based on manual human stimulus annotation guided by heuristics. 
While relatively elementary and limited in several respects, this model has the key practical advantage of being possible to implement on naturalistic driving data of different fidelity across a wide range of scenario types. 

\begin{figure}[t]
	\centering
		\includegraphics[width=0.58\textwidth]{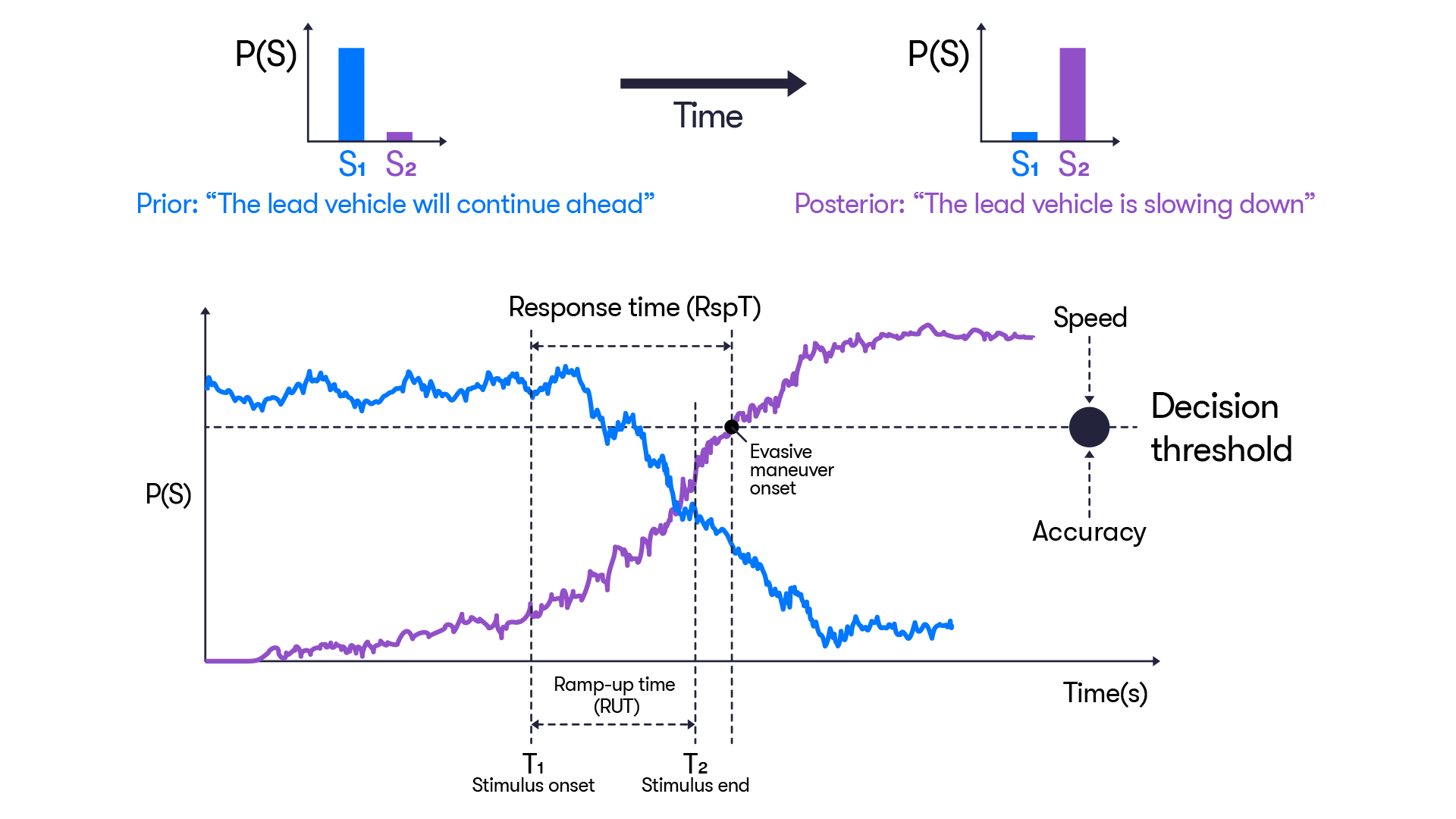}
	  \caption{Conceptual illustration of the time course of a belief update leading to a response, and the key variables for the heuristic model.}
	  \label{heuristic_labels_belief}
\end{figure}

The key idea is to define heuristics to manually annotate a \textit{stimulus onset}, $T_1$, as the first observable \textit{surprising} evidence for the conflict, that is, the first evidence available to the agent that violates the prior belief.
In addition, a second timestamp is annotated that conceptually represents a point where the belief update is complete and the posterior matches the evidence, referred to as the \textit{stimulus end}, $T_2$. 
The difference between $T_1$ and $T_2$ can then be used to obtain an estimate of the rate at which the surprising evidence for the posterior is accumulated, referred to as the \textit{ramp-up time} (RUT). 
By fitting a statistical model with RUT as the independent variable, the response time (RspT) measured from the annotated stimulus onset $T_1$ can be predicted in a novel conflict scenario based on the annotated RUT. 
Figure~\ref{heuristic_labels_belief} illustrates these variables and how they relate to the conceptual model outlined in the previous section.

In practice, the proposed methodology amounts to explicitly defining the subject vehicle (SV) driver's prior belief about the behavior of the other agent for the scenario type in question, referred to as the initial hypothesis ($H_i$), as well as the surprising behavior, referred to as the alternative hypothesis, $H_a$ (posterior belief). 
Since the event to be evaluated is always a traffic conflict (i.e., an unusual, unexpected, event that road users aim to avoid), determining $H_i$ is typically straightforward as it corresponds to the ``normal'', non-conflict, behavior in that situation (e.g., ``the other vehicle will yield'' in Figure~\ref{gradual_scenario} and ``the lead vehicle will continue ahead'' in Figures~\ref{formal_belief_updating}-\ref{heuristic_labels_belief}) and $H_a$ corresponds to the belief that a conflict is impending (e.g., ``the other vehicle cuts in front of me'' in Figure~\ref{gradual_scenario} or the lead vehicle brakes in Figures~\ref{formal_belief_updating}-\ref{heuristic_labels_belief}).
Based on the annotation heuristics defined for the scenario in question (exemplified below),  $T_1$,  $T_2$, and the start of the evasive maneuver (EM) are annotated in naturalistic human crash and near-crash data. 
These annotations are then used to fit a linear model of the response time (RspT = EM - $T_1$) as a function of RUT.
Since EM is annotated as the start of the evasive maneuver, the movement time, the time to move the foot from the accelerator pedal to the brake pedal, is absorbed by the RspT (in line with the SAE J2944 definition of response time, SAE (2015)). 
In addition, since the response onset is here operationalized based on the deceleration signal (described in the next section), any time delay between the brake pedal depression and the deceleration onset (e.g., due to the mechanical delays in braking system) was also included in the RspT.
The fitted RspT model can then be used to generate predictions of response times in new (e.g., simulated) events, as described below.

%% file: 3_methods.tex
\setcounter{figure}{0}

The heuristic model was applied to the modeling of response timing in rear-end conflicts in the SHRP2 naturalistic dataset. 
This section describes the methods used to obtain, select, and annotate the naturalistic driving data. 
While the present analysis only included rear-end conflicts, the approach is straightforward to generalize to other conflict scenario types as exemplified in Appendix~\ref{appendixC_generalizations}.

\subsection{Rear-end conflict scenario definition}

\noindent
The rear end conflict scenarios included in the current data were divided into three main types:
\begin{itemize}
\item S.1: The lead vehicle (LV) brakes surprisingly with the subject vehicle (SV) following behind
\item S.2: The LV moves to exit the lane and then brakes surprisingly with the SV following behind
\item S.3: The LV is stopped or slowing with the SV closing in from behind
\end{itemize}

\noindent
The detailed descriptions of each scenario, including some scenario types that did not show up in the present data set, can be found in Appendix~\ref{appendix_A}.
In the data used for the present analysis, the SV was always a passenger car and the lead vehicle was a motorized road vehicle such as a passenger car (sedan, SUV), minivan, pickup truck or a larger truck.

\subsection{Naturalistic driving data}

The naturalistic conflicts used in the present analysis were obtained from the SHRP2 (Strategic Highway Research Program 2) which is the largest public naturalistic driving dataset available to date and includes data from over 3000 participants, maintained by the Virginia Tech Transportation Institute (VTTI)~\cite{hankey2016description}.
The study participants, between 16 and 80 years of age, drove instrumented vehicles over an extended period of time (up to one year) as part of their daily lives.
Data was collected from six sites across the US: Seattle, WA; Bloomington, IN; Buffalo, NY; State College, PA; Durham, NC and Tampa, FL. 
The data set includes continuous driving data for about 40 million vehicle miles and event data from about 2000 crashes and 7000 near-crashes. 

The data used for the present analysis included event data from rear-end crashes (excluding low-severity, Level 4, crashes) and near crashes with associated video (forward view only, 30 Hz), time series data from onboard sensors (10 Hz), manually annotated eyeglance data and event-level annotations.

\subsection{Visual looming computation}



As described below, the stimulus annotation $T_1$ and $T_2$ in the rear-end conflicts partly relies on visual looming. 
Thus, the looming signal had to be extracted from the data. 
As mentioned above, looming is defined as the expansion rate (time derivative) of the optical angle ($\theta$) subtended by the lead vehicle on the following drivers' retina, $\theta$-dot (\cite{lee1976theory}. 
$\theta$ was extracted directly from the SHRP2 forward video by means of a semi-manual procedure based on computer vision. 
In this method, the visual angle was directly computed based on the measured width of the lead vehicle in each video frame and the camera optics.

To record the width of the lead vehicle in each frame of the video, we developed a computer vision solution for tracking the left and right corners of the vehicle and using the distance between the two as a measure of the optical vehicle width (see Appendix~\ref{appendixB} for further details). 
This method proved to be robust to the strongly varying quality of the SHRP2 front videos, even in turns and night time conditions. 
However, the method does have some limitations. 
For example, the presence of windshield wipers, or a red flickering turn signal light for night time videos could impede the tracking process. 
Still, the method worked reliably in the large majority of SHRP2 events.

The lead vehicle's width in the image, estimated from the computer vision algorithm in pixels (w), was converted into mm using a conversion factor of 720/500 based on the effective number of pixels of the SHRP2 camera and the resized pixel width used by the computer vision algorithm. 
The optical angle $\theta$ was estimated from the camera optics based on the following formula

\begin{equation}
    \theta = 2 \cdot \mbox{arctan}(\frac{w/2}{c_{fl}})
\end{equation}

\begin{figure}[t]
	\centering
		\includegraphics[width=0.49\textwidth]{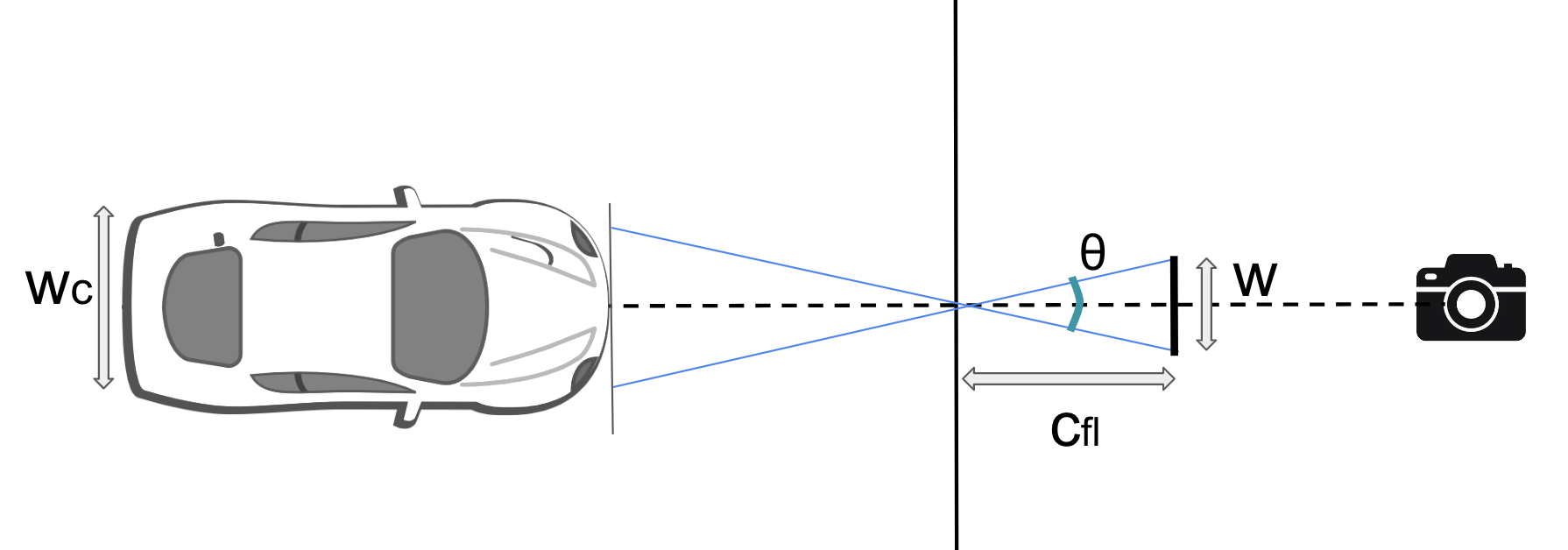}
	  \caption{Calculation of visual angle based on the camera optics.}
	  \label{visual_angle}
\end{figure}

\noindent
where w is the lead vehicle's width in the camera image and $C_{fl}$ the focal length of the camera from the SHRP2 forward video view which was 3.6mm (Figure~\ref{visual_angle}).

\subsection{Data annotation}

For each scenario type S.1-S.3 defined in Section 3.1, the initial hypothesis (prior belief) $H_i$ and the alternative hypothesis (posterior belief) $H_a$ were defined and a set of heuristics developed to guide the manual video data annotation of the stimulus onset ($T_1$) and stimulus end ($T_2$).
Below these heuristics are defined for the most common scenario (S.1) as an example. 
A more detailed definition of the S.1 heuristics and the other rear end scenarios of relevance for the current analysis can be found in Appendix~\ref{appendix_A}.

\textbf{S.1 General description}: The subject vehicle (SV) is following the lead vehicle (LV) at constant speed or slowing down with constant acceleration, when the LV starts decelerating unexpectedly. 
This includes situations where the LV changes lane ahead before decelerating and stop-and-go situations where the SV and LV are initially proceeding slowly with intermittent stops.
$H_i$ hypothesizes that the LV will continue ahead at constant speed or constant acceleration, and 
$H_a$ hypothesizes that the LV brakes and an imminent evasive maneuver is required from the SV to avoid collision.
$T_1$ is determined as what occurs first of (a) the first surprising brake light onset, and (b) the first surprising LV deceleration visible to the SV driver.

Thus, importantly, only surprising brake light onsets count as a stimulus onset. 
In order to determine whether a brake light was surprising in a given situation, a number of additional heuristics were defined, as further described in Appendix~\ref{appendix_A}.
An example of a surprising brake light onset would be a lead vehicle on a freeway braking with no visible traffic ahead. 
An example of an unsurprising brake light onset would be a lead vehicle slowing down to visible traffic queue or red traffic light ahead (cf., Figure~\ref{formal_belief_updating}).
A relatively common situation is also that  the brake light has been on for some time (e.g., due to the lead vehicle going down a hill) in which case there may be no brake light onset at all associated with the current situation. In the case of an unsurprising or lacking brake light onset, $T_1$ was determined as the first visible surprising LV deceleration, which was judged based on the SHRP2 forward video combined with the looming signal, based on the guidelines described in Appendix~\ref{appendix_A}.\footnote{
Due to the lack of precise kinematic information on other road users in the SHRP2 data, this was the best option available. 
In other datasets with more precise kinematic information, other signals, such as the LV's deceleration profile, could be used to determine the deceleration onset based on the same principle.
This shows an example of how the framework (i.e., the heuristic method outlined here), can be flexibly applied to data of different fidelity.
}

$T_2$ is determined as (a) the first point in time at or after $T_1$ where the angular expansion rate (looming) reaches a threshold of 0.05 rad/s. $T_2$ equals $T_1$ if the looming already reached 0.05 rad/s at $T_1$.
This $T_2$ criterion is intended to represent a point where it has become certain to the ego driver that the lead vehicle is closing, according to the general $T_2$ definition above. 
However, the exact value is somewhat arbitrary and the value of 0.05 rad/s was set based on a tradeoff between maximizing the fit of the linear model and not losing data due to the the looming signal never crossing the threshold (e.g., due to the lead vehicle interrupting their braking early). 

$T_1$ and $T_2$ were then manually annotated based on the SHRP2 forward video and the looming trace obtained from the computer vision algorithm described above.
Two data reductionists performed the annotation in parallel. 
If the disagreement between a $T_1$ or $T_2$ annotation was larger or equal to 300 ms, the annotation was resolved in discussion between the two reductionists. 
In the case of a disagreement smaller than 300ms, a final value was determined by the first reductionist considering also the annotation of the second reductionist.

If the SV driver responded to the stimulus onset event before looming reached 0.05 rad/s (and the looming, and the resulting $T_2$, was thus potentially affected by the response), the response was ``removed'' by extrapolating the looming signal from the response onset point and $T_2$ was instead assigned as the first point when the extrapolated signal exceeded the $T_2$ threshold. The extrapolation was conducted by fitting a second-order polynomial to the looming data in a time window from 1s before the stimulus onset until the response onset.
This was done for five out of 36 events (see below) in the present dataset.

The evasive maneuver start (EM) and the eyes-off-path indicator (representing whether the event should be categorized as eyes-on or eyes-off path) were manually annotated based on time series plots overlaying the visual behavior data with kinematic time series data. 
These plots were generated by combining the SHRP2 time series and eyeglance data. 
Since data on pedal and steering wheel inputs are not reliably present in the SHRP2 data, the evasive maneuver start (EM) was manually annotated based on the acceleration time series plots. 
The EM was annotated as the first point in time after the subject reaction onset annotated in the SHRP2 data (representing the SV driver's first visible physical response to the event) where a braking or steering response to the stimulus onset ($T_1$) could be observed.
This involved identifying the start of the deceleration ramp-up that was determined to be associated with the response to the conflict. 
Thus, if for example, the SV driver had already begun slowing down prior to the conflict at a roughly constant low deceleration (for example in anticipation of an intersection), the EM was annotated at the point where the driver started braking harder and the deceleration started ramping up (given that this occurred after the VTTI-annotated subjective reaction). 
This method involved some degree of annotator judgment but was chosen since it is very challenging to programmatically operationalize the evasive maneuver start solely based on acceleration and/or jerk thresholds in a way that applies across all types of situations.  

The binary variable eyes-off-path was set to \texttt{TRUE} if the VTTI eyeglance coding indicated a glance off the forward roadway (where the forward roadway includes the forward, left and right windshield areas of interest) that overlapped at any point with the time interval between T1 and the (VTTI-annotated) subjective reaction onset. 
Since the rear-end conflicts included in the present analysis typically occur in the forward path, the eyes-on-path annotation is here used to operationalize eyes-on-conflict in the NIEON (Non-Impaired Eyes ON conflict) reference model.
The VTTI eyeglance coding also includes eye-closures, which were here included in the eyes-off-path category.
Thus, the eyes-off-path variable at least to some extent accounts for sleepiness-related impairments although it does not cover other types of impairments (e.g., substance-induced) beyond their effects on visual behavior.
Sleepiness and substance-induced impairments were also checked for in the ``Driver Impairments'' variable in the SHRP2 event annotations, as further described below. 
All events were thus manually annotated with $T_1$, $T_2$, EM, and eyes-off-path.

\subsection{Event sampling}

An initial data set of rear-end crashes and near crashes was established by filtering on \texttt{incident\_type} = ``Rear-end, striking'' in the SHRP2 Event Data table. 
This resulted in a set of 119 crashes and 3,669 near crashes. 
To select a manageable subset of events for the present analysis, four selection criteria were initially applied:

\setcounter{figure}{0} 

\begin{figure}[t]
	\centering
		\includegraphics[width=0.45\textwidth]{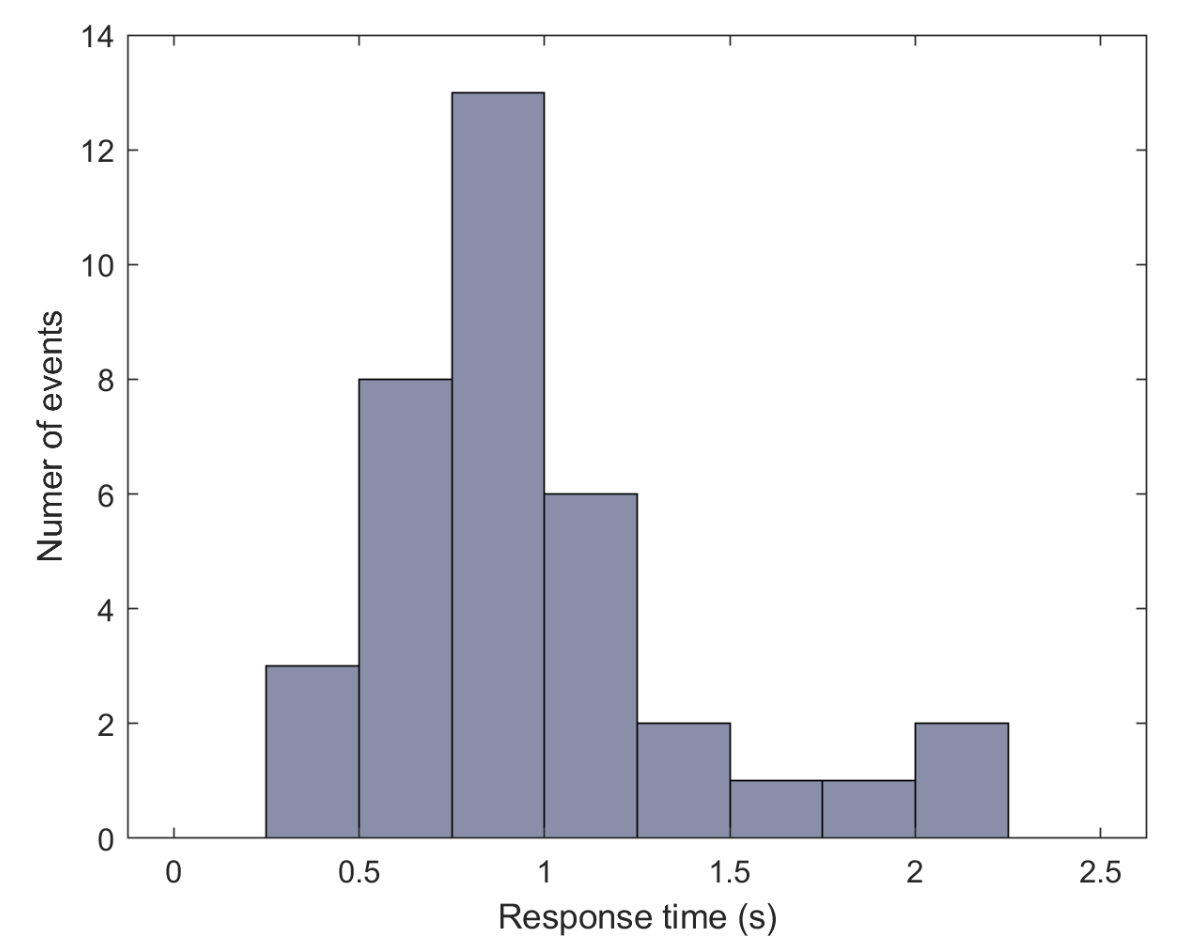}
	  \caption{Distribution of response time (s) for n = 36 lead vehicle braking events.}\label{histogram}
\end{figure}
 
\begin{enumerate}
    \item \textit{Traffic conflict}: whether the event counted as a true traffic conflict was heuristically defined based on whether the vehicle had to brake ``harder than usual'' at (or shortly after) the stimulus onset to avoid collision. This criterion is clearly rather crude (due to the lack of precise kinematic data) and served the purpose of excluding clear cases of non-conflict events. It also protected, at least partly, against selection bias (i.e., when initially benign events became crashes or near crashes because drivers failed to take precautionary action or responded late, e.g., due to eyes-off-path) 
    \item \textit{Stimulus onset visibility}: The stimulus onset ($T_1$) occurred within the forward camera's field of view and was not obscured by fog or darkness.
    \item \textit{Video exists and is of adequate quality}: The video footage was of the required quality to determine $T_1$ and run the looming computer vision algorithm.
    \item \textit{True rear-end scenario}: If the LV performed a lane change away from the SV's lane or a turn, the maximum lane excursion was less than 50\% (i.e., less than 50\% of the SV's 2D-projection on the road left the original lane).
\end{enumerate}

These criteria were applied on randomly selected events from the initial set until four crashes and one hundred near crashes were obtained. 
Crashes and near crashes were sampled separately to obtain a crash/near-crash ratio on the same order of magnitude as in the original SHRP2 dataset. 
Of these 104 events, a total of 60  events (4 crashes and 56 near crashes), were triaged.
The reason for stopping short of triaging all selected events was that the manual data selection and annotation was labor intensive and it was determined that this number of events was sufficient for present purposes. 
One crash and 4 near crashes were excluded during triage due to no driver response.

Since our aim for the present application of the model is to establish a human response timing benchmark based on the NIEON (Non-Impaired Eyes ON conflict) criterion described above, only events classified as \texttt{eyes-off-path = FALSE} and with non-impaired drivers were retained from the triaged events.    
Impairments were identified using the  ``Driver Impairments'' variable in the SHRP2 event table, where we define ``impairment'' as driver states related to sleepiness or substance-induced impairments related to alcohol or other drugs (note, the SHRP2 Driver Impairments variable also include emotional states such as ``anger'').
Fourteen eyes-off-path events and one event annotated as ``Drowsy, sleepy, asleep, fatigued'' were excluded from the triaged dataset. 
In addition, 3 more events were excluded due to missing looming tracking, and one more was removed due to triage error, resulting in a total of 36 events remaining for analysis.
Of these twenty-five were of type S.1, four of type S.2 and seven of type S.3.

%% file: 4_results.tex
\setcounter{figure}{0} 

\subsection{Response time distribution and response model fitting}

\begin{figure}[t]
	\centering
		\includegraphics[width=0.5\textwidth]{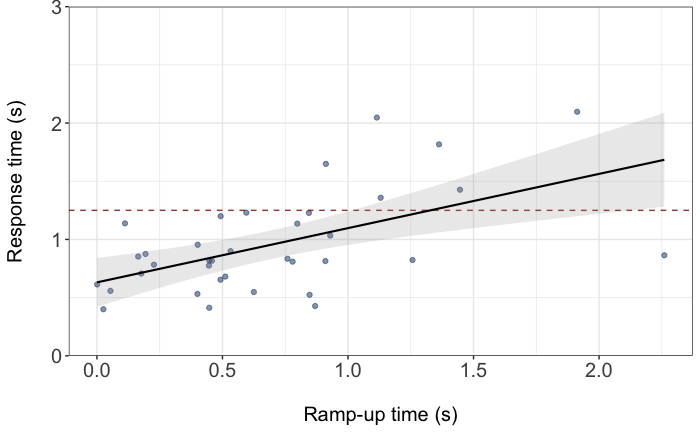}
	  \caption{Ramp-up time (RUT) versus response time for (n=36) naturalistic rear-end crashes and near crashes from the SHRP2 dataset (shown as points).
	  The black line represents the linear model fitted to the data with k=0.47 and m=0.63 and the gray ribbon represents the 95\% confidence interval. 
	  The red dotted line represents the canonical response time of 1.25 sec for rear end conflicts proposed by Green (2000), demonstrating how a ``flat'', canonical response time overestimates response times for short RUTs, and underestimates response times for long RUTs.}
    \label{regression_result}
\end{figure}

The distribution of response times in the rear-end conflicts is shown in Figure~\ref{histogram}. 
As is typical for response time distributions, the data is positively skewed and exhibits a heavy tail.
However, based on the framework outlined above, response timing is expected to depend on the situation and more specifically, on the rate of surprising evidence here represented by ramp-up time (RUT). 
As expected and in line with existing literature, response times tend to increase with ramp-up time with a Pearson product-moment correlation of 0.55  (Figure~\ref{regression_result}).  

\begin{figure}[t]
	\centering
		\includegraphics[width=0.45\textwidth]{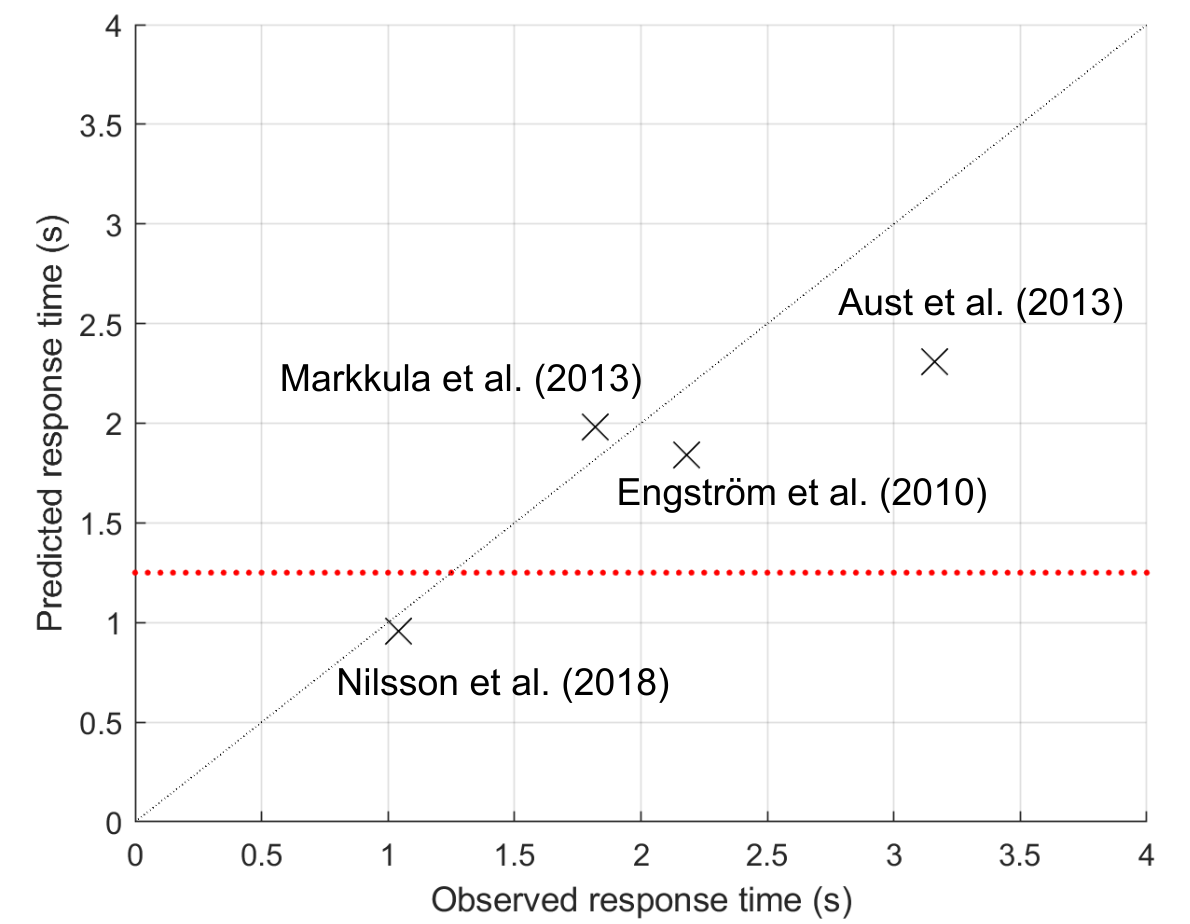}
	  \caption{Response time predicted by the model vs. average response time observed in the studies. 
	  The plot also includes the prediction implied by~\cite{green2000long} proposed canonical, situation-independent response time for brake lights.}\label{validation}
\end{figure}

We fit a linear model, $RspT = m + k \cdot RUT$, to capture the relationship between ramp-up time and response time.
As will be further addressed in the Discussion section, there are more sophisticated ways to model the data, which, for example, would account for the heteroscedasticity that can be observed in Figure~\ref{regression_result} (i.e., longer ramp-up-times show more variability in response times than shorter ramp-up-times). 
The linear model was chosen here for simplicity and interpretability. 
The linear regression yielded model parameter values of k=0.47 and m=0.63 (p<0.001) thus resulting in the following model:

\begin{equation}
    \mbox{RspT} = 0.63 + 0.47 \cdot \mbox{RUT}
\end{equation}

The resulting fit and 95\% confidence interval is shown in Figure~\ref{regression_result}. 
For an instantaneous ramp-up of surprising evidence (RUT=0s), the model predicts a response time of 0.63 seconds, and for every one second increase in ramp-up time, the response time is expected to increase by 0.47s. 

Figure~\ref{regression_result} also illustrates the canonical response time of 1.25s suggested by~\cite{green2000long} for responses to ``normal, but common, signals such as brake lights'' (p. 213).~\footnote{It should be noted that Green has since also pointed out this limitation, as well as other challenges in defining and measuring perception-response time. See Green (2009) and https://www.visualexpert.com/Resources/realprt.html for excellent discussions.}
As can be seen, since such a model does not account for the situatedness of response time (represented by the RUT in our model), it would overestimate the response time in fast-developing scenarios (short RUT) and underestimate response time in scenarios where the evidence comes in at a slower rate (long RUT). 
As shown below, the same issue applies for response times reported in published driving simulator studies.

\subsection{Validation against experimental response time data}

To further validate the model, we compared the model response time predictions to the observed average response times from  four existing driving simulator rear-end conflict studies (\cite{aust2013effects,markkula2014modeling,nilsson2018effects,engstrom2010effects}).

As can be seen in Table 1 and Figure 14, the observed average response times differ widely between the published studies.
Based on the current framework (and previously discussed in~\cite{markkula2016farewell}), this can be largely attributed to differences in scenario kinematics, in particular the initial time gap and the lead vehicle deceleration rate (see Table 1).
The study with the shortest initial time gap and the hardest braking lead vehicle~\cite{nilsson2018effects}  reported the shortest average response time (1.04s) and studies with longer initial time gap and/or a softer braking had longer response times (e.g.,~\cite{aust2013effects}, 3.16s, a difference of more than 2s).
This is in line with other empirical studies and meta-analyses reviewed above (\cite{li2014effectiveness,engstrom2010scenario,engstrom2018simulating,markkula2016farewell,bianchi2020drivers}). 
Existing modeling work (\cite{engstrom2018simulating,markkula2016farewell} has shown that the observed response times can be predicted with a relatively high accuracy by models based on visual looming models (since visual looming depends directly on kinematic variables such as initial time headway and deceleration rate). 
Of particular relevance for the present analysis,~\cite{markkula2016farewell} analyzed the same four studies included here and found that a simple response model based on inverse $\tau$ (where $\tau$ is the optically specified time-to-collision defined as $\theta$ / $\theta$-dot), fit to SHRP2 data, accounted for 73\% of the variance in the average response times.
Thus, it is interesting to examine to what extent our model is also able to predict these unseen experimentally obtained response times.

As explained above, in our heuristic model the kinematics dependence is accounted for by the RUT, which, for rear-end conflicts, is largely determined by the looming threshold ($\theta$-dot, or optical expansion rate > 0.05 rad/s). 
Thus, we extracted the ramp-up times from the existing studies using the same stimulus annotation heuristics that was used above for the SHRP2 data. 
These RUTs were then input to our model to generate predictions for the average response times reported in the studies.
The simulator studies all included an initial truly surprising event followed by events that were repeated for the same subject and thus somewhat expected. Only data from the first, truly surprising events was used in the current analysis. 

The scenarios in all four experimental studies were of the type S.1 defined above (the LV brakes surprisingly with the SV following behind).
As defined above, $T_1$ was the surprising brake light onset and $T_2$ the point where looming exceeded 0.05 rad/s. 
The looming signal was computed based on the scenario kinematics reported in the respective papers, as further described in~\cite{engstrom2018great}. 

As can be seen in Figure~\ref{validation}, our model predicts the observed response times from the published simulator studies relatively well, albeit not perfectly. 
This is despite the fact that three of the four simulator studies had RUTs outside the range of the data used to fit the model (where the maximum RUT was 2.2 sec).
The model fit was evaluated in terms of $R^2$, 

\begin{equation}
    R^2 = 1 - \frac{\sum_{k=1}^{n}(p_k - o_k)^2}{\sum_{k=1}^{n}(o_k - \bar{o})^2}
\end{equation}

\noindent
where k = 1, ..., N; N=4, enumerates the four studies, $o_k$ are the observed average response times, $p_k$ are the average response times predicted by the model and $\bar{o}$ is the average across the four observed average response times. 
The $R^2$ value for the model was 0.62, meaning that the model explains 62\% of the variance in the average response time from the four studies.
This is slightly lower, but comparable to $R^2 = 73\%$ obtained by~\cite{markkula2016farewell}.
 
These results demonstrate (1) that human response times are strongly situation-dependent (in line with previous studies), and (2) that our simple heuristic model captures this dependence and generalizes relatively well to unseen response time data even outside the range in which it was fit.  

The results also further demonstrate the inadequacy of situation-independent response time models. Figure~\ref{validation} includes the canonical response time of 1.25s suggested by Green (2000) discussed above, which was also included in Figure 13 above. This situation-independent (flat) prediction is relatively close to the 1.04 s observed in Nilsson et al. (2018) but would considerably underestimate the other observed response time values (and as shown above, it substantially overestimates the RspTs for shorter RUTs in the SHRP2 data). 
In contrast, the current model, accounting for scenario dependence in terms of RUT, predicts all four observed average response time values relatively well.

\begin{table*}[t]
\caption{Kinematics of the lead vehicle (LV) braking scenarios in the four simulator studies, detailing: 
the number of observations (N), the subject vehicle (SV) type, 
the instructed speed ($v_{V1}$) and the LV speed at the onset of deceleration $v_{LV}$, (both in kph),
the time gap at the start of the LV's breaking (initial time gap),
the lead vehicle deceleration d(g)),
the stimulus onset ($T_1$), end ($T_2$), ramp-up time (RUT), and the average observed and predicted response times, (Obs $\mu_{RspT}$ and Pred $\mu_{RspT}$, respectively), in seconds.}
\begin{tabular*}{\tblwidth}{ccccccccccccc} 
\toprule
 Study & N & SV type & $v_{V1}$ & $v_{LV}$ & Init. time gap &  d(g) & Obs. $\mu_{RspT}$ & $T_1$ & $T_2$ & RUT & Pred. $\mu_{RspT}$\\
\hline
    \cite{engstrom2010effects} & 20 & Car   & 70 & 80 & 1.5 & 0.51 & 2.18 & 0 & 2.6 & 2.6 & 1.87 \\
    \cite{aust2013effects}       & 8  & Car   & 90 & 90 & 2.5 & 0.55 & 3.16 & 0 & 3.6 & 3.6 & 2.36 \\
    \cite{markkula2014modeling}     & 48 & Truck & 80 & 80 & 1.5 & 0.35 & 1.82 & 0 & 2.9 & 2.9 & 2.02 \\
    \cite{nilsson2018effects}     & 10 & Car   & 80 & 48 & 1.3 & 0.60 & 1.04 & 0 & 0.7 & 0.7 & 0.94 \\
\hline
\end{tabular*}
\label{tbl_validation}
\end{table*}

%% file: 5_computational.tex
\setcounter{figure}{0}  

There are several ways to implement the proposed framework computationally. 
This has the practical advantage of eliminating, or at least reducing, the need for manual annotation, which would allow the framework to scale to larger datasets.
It would also make the models mathematically precise, thus removing the subjectivity inherent in the heuristic method described above.
Here we explore a computational model based on the general evidence accumulation framework originally proposed by~\cite{markkula2014modeling} and further developed in subsequent work (\cite{engstrom2018great,dinparastdjadid2020modeling,markkula2016farewell,svard2017quantitative}).
The key novel feature is that the SV driver's prior belief is automatically generated by a machine-learned generative behavior prediction model, which (in principle) makes the model generalizable across scenarios.
The surprising evidence is then computed based on the actual agent behavior relative to the prior belief and input to a traditional evidence accumulation model.
Thus, the quantity that is accumulated is surprising evidence rather than the evidence/stimulus per se.
This is conceptually similar to existing evidence accumulation response models operating on (visual looming) prediction error (\cite{bianchi2020drivers,engstrom2018great,svard2017quantitative}).
However, the predictive models used in these studies were relatively simple and stimulus (looming) specific and, hence, these models were not generalizable beyond the specific scenarios and applications they were developed for (e.g., in~\cite{bianchi2020drivers}, the modeled phenomenon was responses to silent failures of an Adaptive Cruise Control system in rear-end situations). 

Here we generalize the notion of prediction error to the notion of surprise computed based on a generative model (see (\cite{dinparastdjadid2023surprise}) for a detailed account on the computation of surprise in real-world scenarios based on machine-learned generative models).
A generative model is a common concept in statistics and machine learning, and refers to a model that generates statistical predictions about observations. 
These predictions typically take the form of probability distributions, which, in the present framework, represent the prior beliefs of the SV driver about the behavior of other road users. 

\begin{figure}[t]
	\centering
		\includegraphics[width=0.5\textwidth]{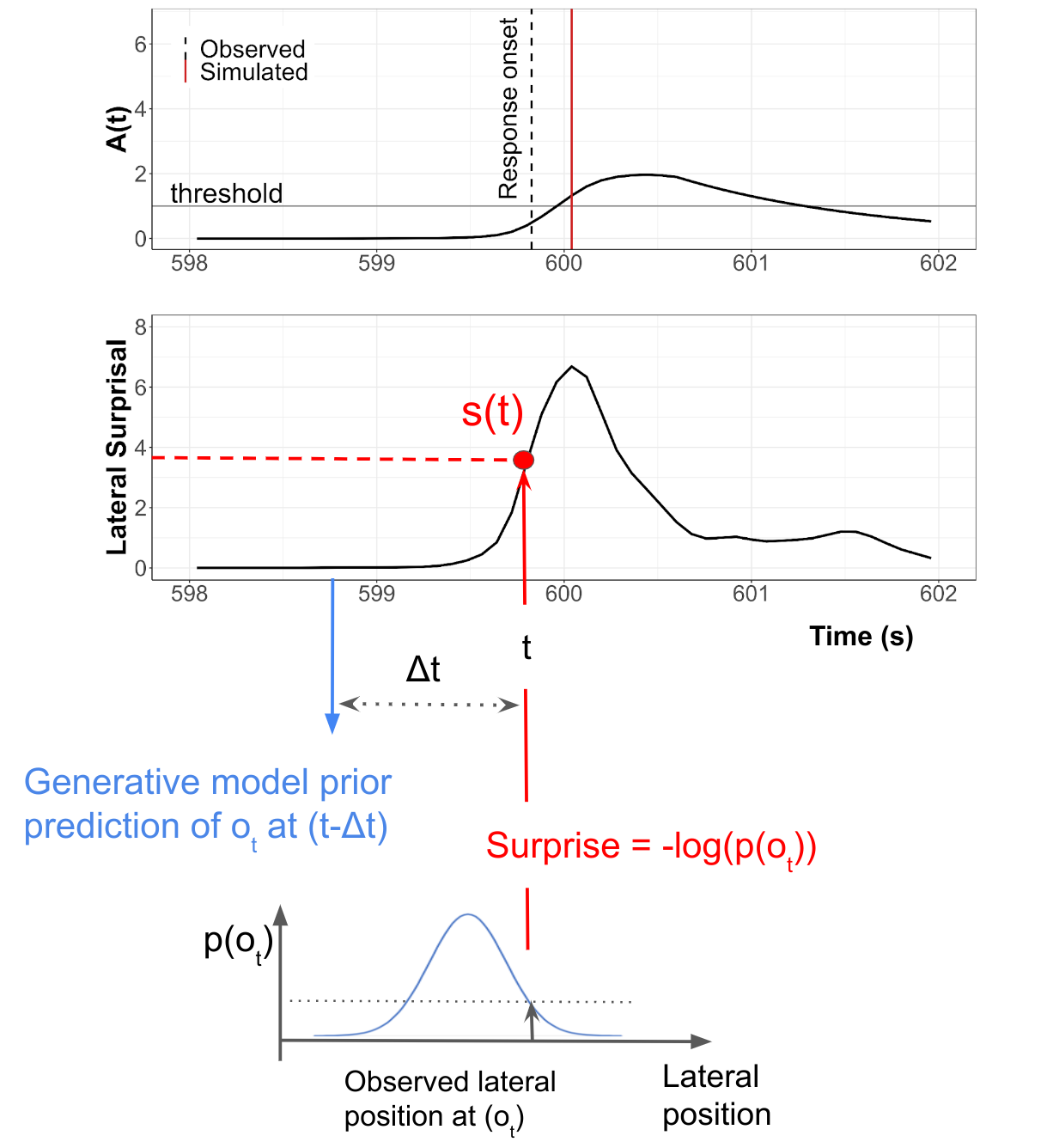}
	  \caption{Demonstration of a computational implementation of the model defined by equation (4).}
	  \label{fig_computational}
\end{figure}

Given a prior belief at a time t-$\Delta$t, represented, for example, by a mixture of Gaussians, the surprise $s_t$ of a specific observation $o_t$ at the current time $t$ can be quantified as $-\mbox{log}(o_t)$, also known as surprisal~\footnote{
Surprisal is also known as Shannon surprise or self information. 
An alternative way to operationalize surprise is in terms of the Kullback-Leibler (KL) divergence between a prior and posterior probability distribution (\cite{itti2009bayesian}).
} (see the graph at the bottom of Figure~\ref{fig_computational}).
This surprise signal can then be used as the input to an evidence accumulation model, similar to how the prediction error was in the studies cited above.

A specific example of such a model is given in Equation~\ref{equation_4} and a simulation of this model applied to a lateral conflict scenario  is shown in Figure~\ref{fig_computational} (top).
The lateral conflict scenario was chosen to demonstrate how this type of approach, based on a general machine-learned generative  model, naturally generalizes across scenario types.
The purpose of including this simulation here is merely to provide a concrete example of how the proposed response timing framework can be implemented computationally. Further exploration of this and similar computational models is left for future publications.  

The specific generative model used to produce the prior belief is known as Multipath and described in~\cite{chai2020}.
After being trained on large quantities of driving data, Multipath generates prior beliefs of the possible future lateral and longitudinal positions of the cutting-in vehicle along a set of predicted trajectories in the form of probability distributions (Gaussian mixtures) at different time steps into the future. 

In the current simulation, another vehicle (the POV) unexpectedly cuts in front of the SV (the responding vehicle) from the adjacent lane thus generating a traffic conflict. 
The key observational variable is thus the lateral position of the POV ($y_t$).
The surprise $s(t)$ was computed as the the negative log likelihood ($-\mbox{log}(P(y_t))$ of the POV's currently observed lateral position $y_t$ under the prior belief (Multipath Gaussian mixture) generated $\Delta$t seconds earlier ($\Delta$t was here set to 1s). 
The surprise signal $s(t)$ was then input into the evidence accumulation model defined in equation~\ref{equation_4}, which is similar to existing evidence accumulation models (e.g.,~\cite{bianchi2020drivers,markkula2016farewell,svard2017quantitative,svard2021computational}).

\begin{equation}
    \label{equation_4}
    \frac{dA(t)}{dt} = k s(t) + \lambda A(t) + v(t)
\end{equation}

where $A(t)$ is the activation which accumulates towards the response threshold $T$ representing a tradeoff between decision speed and accuracy (\cite{bogacz2006physics,dinparastdjadid2020modeling}.
Thus, in terms of the proposed framework, $A(t)$ here conceptually represents the strength of the posterior belief (i.e., the purple trace in Figure~\ref{heuristic_labels_belief}); $k$ is again a constant, $\lambda$ is a leakage term, and $v(t)$ is a noise term (the specific formulation of the model may vary depending on the application). 
The evasive maneuver is assumed to be initiated when the activation $A(t)$ passes the threshold $T$, here set to 1. 
The gain and leakage parameters, and the noise function $v(t)$, were fit based on ground truth response onset data from lateral cut-in conflicts obtained from the heuristic model described above using Approximate Bayesian Computation (ABC) parameter estimation (\cite{dinparastdjadid2020modeling,beaumont2010approximate,toni2009approximate}).

The simulation results are shown in Figure~\ref{fig_computational}. 
Before the other vehicle starts cutting in (at about t=599.4s) the generative model (correctly) predicts that it will remain in the adjacent lane and the surprise is close to zero.
When the cut-in begins this represents an increasingly surprising observation at the tail of the prior belief (i.e., the probability of the currently observed lateral position $o_t=y_t$ of the cutting-in vehicle is low and, thus, the surprise $-log(o_t)$ is high). 
The surprise signal then drives the activation $A(t)$ towards the decision threshold for initiating an evasive response.

%% file: 6_discussion.tex
The key contribution of the present paper is the proposed novel framework for measuring and modeling response times in naturalistic traffic conflicts involving automated driving systems as well as in other traffic safety domains. 
The framework, based on the general notion of predictive processing in cognitive neuroscience (\cite{clark2013whatever,clark2015surfing,clark2023experience,engstrom2018great}), suggests that response timing must be understood relative to the subject's prior belief and is always embedded in the dynamically evolving situation.
Thus, a key distinguishing feature of the framework, which sets it aside from the traditional information processing view illustrated in Figure~\ref{brain_processing} and most existing evidence accumulation models, is that the evaluation of response time always starts with identifying the road user's prior belief in the given situation.
The response is then modeled as driven by perceived violations to the prior belief, that is, surprise. 
The generality of this basic principle makes the framework applicable to any type of traffic conflict scenario, human response data of different fidelity and can be implemented in different ways, from manual annotation based on heuristic rules to full-blown computational computational models, as demonstrated above.

This framework resolves two key limitations with traditional notions of response time when applied in naturalistic scenarios: (1) The situation-dependence of response timing and (2) how to unambiguously define the stimulus.
Resolving these issues is a challenge that must be addressed by any response timing model intended to be applied in naturalistic traffic conflicts.
The first challenge is resolved by representing the response process as a gradual accumulation of evidence in a belief updating process, similar to existing evidence accumulation models of road user response time (e.g.,~\cite{markkula2014modeling}). 
The second challenge is resolved by defining the stimulus as surprising evidence, that is, evidence that violates the initial (prior) belief. 

Resolving the first challenge (i.e., accounting for situation dependency of response timing), is required to accurately represent human response performance across different scenarios, for example with different kinematics. 
As shown in Figures~\ref{regression_result}, \ref{validation}, and Table 1, the variation in response timing due to the scenario-specific factors can be very large (> 2s in the studies listed in Table 1). 
Traditional models, assuming a fixed, situation-independent response time across events, would overestimate the human response time for fast developing events (short RUT) and underestimate the response time for more gradually developing (long RUT) events.
This fundamental issue with traditional perception-response time models was illustrated by including the ``flat'' prediction from~\cite{green2000long} proposed canonical response time for rear-end events in Figures~\ref{regression_result}, \ref{validation}. 
Specifically, in Figure~\ref{regression_result}, it was shown that the canonical response time overestimates response times for RUT close to zero with about 100\% (i.e., the response time predicted by the canonical model is about twice the observed average response time), but at the same time underestimates the response times in the driving simulator scenarios with long RUTs by several seconds.
It should be emphasized that this problem is not specific to \cite{green2000long}'s canonical model, but holds for any traditional (PRT-based) scenario-independent response-time model (as already mentioned, Green has, in more recent work (e.g.~\cite{green2009perception}), provided excellent discussions on these and other challenges with applying traditional PRT in the traffic domain).
Clearly, such a model is not suitable as a basis for representing human response performance in naturalistic traffic conflicts, for example with the purpose to establish human performance benchmarks.
In contrast, the framework and model proposed here explicitly accounts for this scenario dependency and is able to account for response times across a wide range of scenarios with different kinematics.

Resolving the second challenge by defining stimulus onset in terms of surprise onset yields a principled way to operationalize the stimulus that road users actually respond to in naturalistic conflict situations, where the stimulus is not pre-defined by the scenario and/or experimental instructions. 
For example, if the brake light onset is arbitrarily defined as the stimulus in a naturalistic scenario, but the brake light onset was already expected by the following driver, the driver may not respond until much later (e.g. to surprising looming).
This would lead to very long measured ``responses times'' which no longer represent responses to the actual conflict. A common example  of this type is scenario S.2 in Appendix~\ref{appendix_A} (discussed in~\cite{engstrom2018great}), where the lead vehicle is slowing down to exit the road and the following driver intentionally closes in on the lead vehicle, thus producing significant looming, based on the expectation that the lead vehicle will complete the turn without stopping. 
The conflict is then initiated by the lead vehicle stopping unexpectedly (e.g., due to an obstacle ahead). 
A response model based on the accumulation of looming per se would not accurately capture the human response behavior since the perceived looming was initially intentional, thus unsurprising, and does not drive a response from the following driver. 
Rather, the true stimulus onset to which the following driver reacts is the surprising lead vehicle stopping which produces stronger looming cues than expected. 
As already mentioned,  this phenomenon has been addressed, in the specific context of rear-end conflicts, by existing accumulation models operating on looming prediction error (conceptualized in~\cite{engstrom2018great} and implemented in~\cite{svard2017quantitative,bianchi2020drivers}). 
The key novelty of the current formulation in terms of surprise computed relative to a prior belief (which may be manually annotated or implemented computationally by a generative model) is that it generalizes this idea to any possible conflict scenario, as further discussed below.

The present paper mainly focused on demonstrating how the framework can be implemented in a practically feasible way, by means of a heuristic model based on manual stimulus annotation. 
It was shown how this relatively simple model, when applied to naturalistic crashes and near crashes from the SHRP2 dataset, was able to account for human response timing in a variety of naturalistic rear-end conflicts in the SHRP2 data.
The model fit to SHRP2 data made reasonable predictions of average observed response times in four driving simulator studies (Figure~\ref{validation}).

While the current analysis focused on rear-end scenarios, it is in principle straightforward to generalize the current heuristic model to other types of conflict scenarios, as exemplified in Appendix~\ref{appendixC_generalizations}. 
Specifically, applying the model to a novel scenario amounts to (1) defining the initial hypothesis (prior belief) and the alternative hypothesis (posterior belief), and (2) defining the stimulus onset ($T_1$) and end ($T_2$) for that particular scenario.
In practice there will always be edge cases that defy a clear-cut classification into a set of predefined model scenarios but, in our experience, a set of about thirty model scenarios covers the majority of real-world events encountered. 
Validating the heuristic model on a wider range of conflict scenario types is a topic for future work. 
The computational model and simulation example in Figure~\ref{fig_computational} applied to a lateral cut-in scenario further illustrates how models based on the present framework can be generalized across scenarios. 

It should be noted  that the current, relatively simplistic, heuristic implementation of the framework is subject to several limitations.
First, the heuristics for manually annotating $T_1$ and $T_2$ are limited in their precision. 
Using specific physical features such as (surprising) braking onsets, brake lights, looming thresholds or lane boundary crossings as the basis for $T_1$/$T_2$ annotation has the advantage of ensuring consistency in human annotations, but may not perfectly represent the surprising stimulus that the road user reacts to.
The manual annotation process itself is also susceptible to some degree of subjectivity. 
In addition, in the present study, the evasive maneuver onset annotation based on the acceleration time series signal also relied to some extent on annotator judgment in determining the first response to the conflict (without direct access to reliable data on pedal or steering wheel input).
Thus, the relative simplicity of this heuristic approach is likely responsible for at least a portion of the variance observed in the response time data which is not explained by RUT. 

Second, the current heuristic model assumes that the driver responds to a conflict with a distinct single evasive (braking and/or steering) response  which is triggered after the modeled response time. 
This assumption typically holds in truly urgent situations like the SHRP2 crashes and near crashes (~\cite{markkula2016farewell}) where the driver needs to initiate an immediate evasive maneuver to avoid or mitigate a collision. 
However, it breaks down in less critical situations where drivers may choose to respond in a more gradual or stepwise fashion, for example, initially slowing down before braking harder when the hazard is realized. 
For this reason, the current heuristic model only applies to traffic conflicts which, by definition, require an imminent response. 
However, this leads to the further limitation that the model is unable to account for overt (e.g., slowing down, moving the foot to the brake pedal) or covert (e.g., increased attention) responses to (surprising) pre-conflict cues.
However, while this is a limitation of the current heuristic model implementation, it is not a limitation of the framework in general. For example,~\cite{svard2017quantitative,svard2021computational} present a surprise-based (prediction error) evidence accumulation model that performs intermittent braking adjustments to cancel looming prediction error in a stepwise fashion, which fits perfectly with the general framework proposed here.
However, adding these types of features comes with the cost of increased model complexity and, possibly, reduced generalizability across scenario types.   

Third, there are numerous factors known to influence human response time (reviewed e.g.,~\cite{olson1989driver,green2000long}) that are not accounted for by the current model.
This includes driver states (other than eyes on/off road such as intoxication, fatigue and cognitive load), individual characteristics (e.g., age, driving experience) and visibility conditions (darkness, fog, rain etc.). 
The inclusion of such factors in the model is an interesting area for further work. 
However, in practical applications some or all of these factors may not be known and hence, cannot be used as covariates in real-world ADS applications for predictive purposes.
In this case, the variance associated with these factors may could be captured by more advanced statistical models.  

Fourth, the number of data points used to fit the model was relatively limited in the current study.
As mentioned above, the SHRP2 dataset contains over 3500 rear-end near crashes and crashes.
However, due to the labor intensive semi-manual data preparation and annotation procedures required for applying the heuristic model, only a small subset of the available events was included in the present analysis.
An automated model (such as that illustrated in Section~\ref{sec5_computational}) would thus enable the model to be fit to much larger datasets than is currently possible with the heuristic model.      

Finally, the ability of any response timing model to truly represent an ``attentive and non-impaired driver'' benchmark is limited by the data used to fit it. 
In the current study, NIEON (Non-Impaired Eyes ON conflict) was operationally defined by (1) gaze being directed through the windshield toward the forward path (based on the manually annotated SHRP2 eyeglance time series annotations) during the conflict and (2) a lack of sleepiness and intoxication-related impairment (based on the manually annotated SHRP2 ``Driver Impairments'' event variable).
Thus, there was no guarantee that the driver's gaze was actually directed towards the conflict, but for the rear-end conflicts currently analyzed it seems like a reasonable assumption that eyes-on-path also implies eyes-on-conflict (which may not be true for other conflict types). 
Moreover, while overt attention can be reliably inferred from visual behavior, covert attention (which may be reduced by daydreaming or cognitive load) is very hard to observe in naturalistic driving data. Also, even though the SHRP2 Driver Impairments variable included categories related to substance induced impairments, this is typically hard to annotate based on video alone. 
Thus, while we did our best to only include events that satisfied the NIEON criterion, it is possible that the data still contained some instances of driver inattention or impairment.

In future implementations of the framework, it would be desirable to eliminate the subjective element, and save labor associated with the manual annotation by replacing the heuristic model with a fully automated computational model, such as that discussed in Section~\ref{sec5_computational}.
This requires automatic generation of prior expectations across a wide range of scenarios as the basis for computing surprise. 
We showed how such a model, based on evidence accumulation and a machine-learned generative behavior prediction model, could be formulated and implemented in simulation. 
This type of model is, in principle, generalizable to any scenario for which the generative model can generate statistical behavioral predictions for other agents.
In addition, the generative model could also account for how drivers with different driving histories (e.g., US vs. French drivers) may generate different prior beliefs in the same situation ,and generate specific predictions on how this would affect their response performance. 
However, achieving the same level of generalizability and robustness as with the current heuristic model is practically challenging for a number of reasons. 
One key challenge is to fit the model to human conflict data.
This requires the generative model to run on naturalistic traffic conflict data (such as SHRP2) that contains detailed information about driver state (e.g. visual behavior and impairments) in real crashes and near crashes. However, this data is typically limited in terms of environment sensors (e.g., the location of other agents in the scene), which is the key input to the generative model.
Recent developments in reconstructing naturalistic human crashes and near crashes in simulation is one promising way forward. 
Furthermore, to facilitate the establishment of more automated human reference models, there is a strong need to establish new sources of publicly available large-scale naturalistic driving data that contain both high-fidelity driver state data (in particular visual behavior) and a detailed representation of surrounding vehicles and the road infrastructure. Recent advances in computer vision (and other sensor) technology makes this a realistic possibility but large scale efforts (similar to SHRP2) are needed to collect the data and make it publicly available.  

The current modeling framework has numerous applications beyond establishing human response timing reference models for ADS. 
For example, the framework may be used as the basis for establishing \textit{distributions} of road user response timing with associated probabilities (see e.g.,~\cite{bargman2015does} for existing work in this area). 
While the current analysis focused on deriving a single reference response time in a given scenario (here representing non-impaired drivers with their eyes on the conflict), probabilistic response time models would account for the full range of agent responses, including rare long-tail response times associated with inattentive or impaired drivers and can be used to estimate crash risk in specific scenarios and predict crash rates at the population level. 

Finally, while the present paper focused exclusively on modeling responses in critical traffic conflicts, the proposed predictive processing / belief updating framework can potentially also be used to model driving behavior in non-conflict situations.
In particular the notion of allostatic control (\cite{sterling2012allostasis,stephan2016allostatic} is useful as the basis for modeling precautionary behaviors whereby human drivers manage to stay out of conflicts in the first place. While responses to (surprising) conflicts is an example of homeostasis (responding to feedback indicating deviations from the expected state), allostasis amounts to anticipatory behavior that generates (unsurprising) evidence that aligns with the prior expectations (e.g., to stay out of conflicts). 
Driver models implementing such active inference (\cite{wei2022modeling,wei2023carfollowing}) may be used to establish reference models for conflict avoidance behavior and is a very interesting avenue for future research.

\section{Acknowledgements}

We would like to thank Dr. Gustav Markkula and Dr. Jonas B\"{a}rgman for valuable feedback that greatly helped to improve the manuscript.

%% file: appendix_A.tex
\section*{Appendix}

\section{Model scenarios}\label{appendix_A}

\setcounter{figure}{0}  

\noindent
This Appendix describes  the model scenarios and stimulus annotation heuristics used in the present study. 

\subsection*{S.1 The LV brakes surprisingly with the SV following behind}

\begin{figure}[h]
	\centering
		\includegraphics[width=0.4\textwidth]{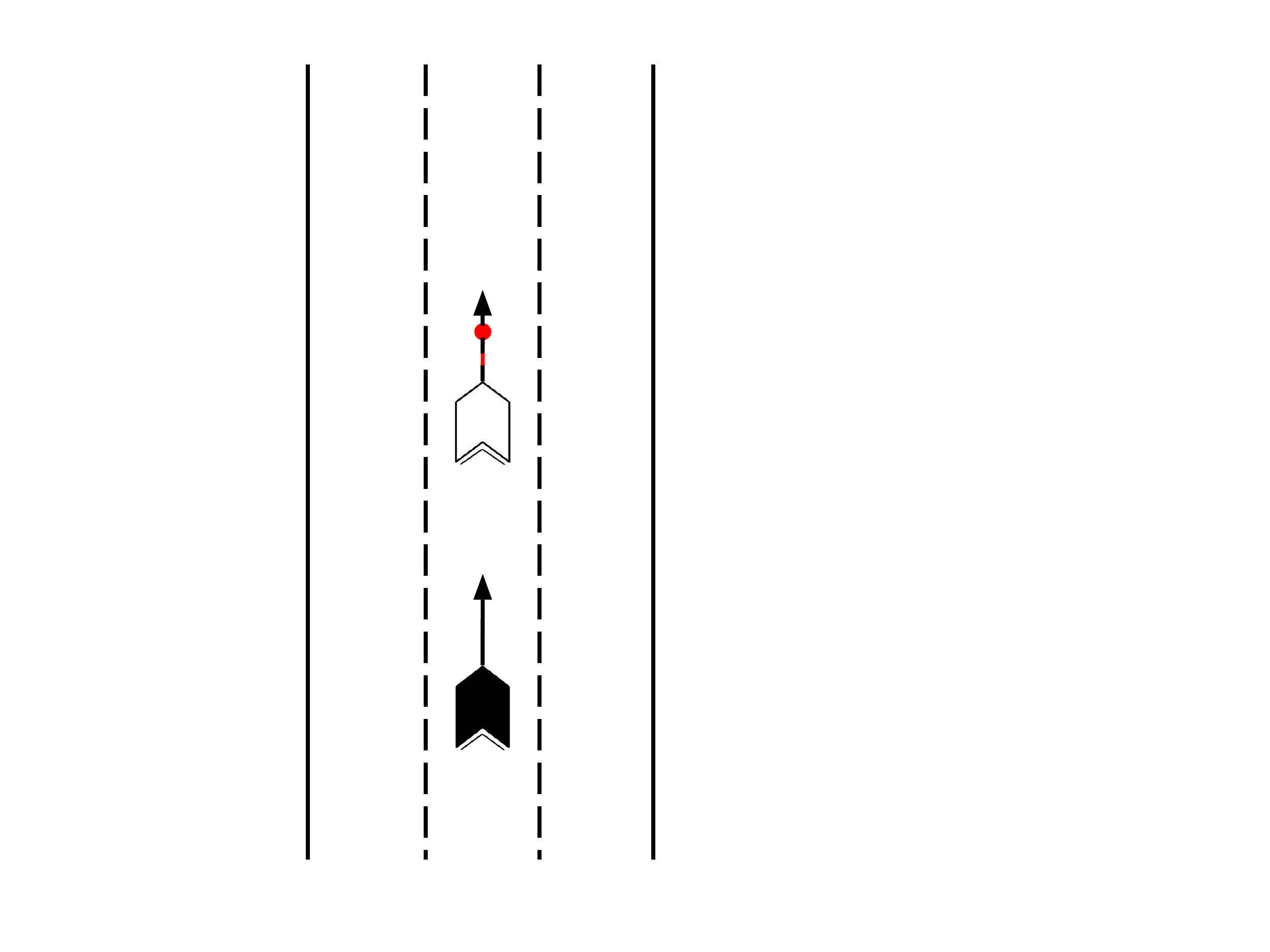}
	  \caption{Pictogram illustrating scenario S.1.}
	  \label{appendix_fig1}
\end{figure}

The SV is following the LV at constant speed, or slowing down at constant acceleration, and the agent surprisingly starts decelerating. This includes situations where the LV changes lane ahead as well as stop-and-go situations where the SV and LV are proceeding slowly with intermittent stops (Appendix Figure~\ref{appendix_fig1}).
\vline

\noindent
Prior belief ($H_i$)
\begin{itemize}
    \item The LV will continue at constant speed or constant acceleration.
\end{itemize}

\noindent
Surprise ($H_a$)
\begin{itemize}
    \item The LV brakes.
\end{itemize}

\noindent
Stimulus onset ($T_1$).
\begin{itemize}
    \item Whatever occurs first: 
    \begin{itemize}
        \item First surprising LV brake light onset.
        \item First surprising LV deceleration visible to the SV  driver.
    \end{itemize}
\end{itemize}

\noindent
Stimulus end ($T_2$)
\begin{itemize}
    \item The first point in time at or after after $T_1$ where the angular expansion rate (looming) of the LV reaches 0.05 rad/s.
    \item  If the looming already reached 0.05 rad/s at $T_1$, or if it never reaches 0.05 rad/s (e.g. due to early braking or swerving), $T_2$ should be set to $T_1$.
\end{itemize}

\noindent
NOTES

1. \textit{Surprising brake light onset}: A brake light onset is unsurprising if it occurs in a situation where the LV is expected to tap the brakes and/or slow down. Examples of unsurprising brake light onsets / agent deceleration events include:
\begin{itemize}
    \item The LV and the SV approach an intersection that is visible to the SV  driver.
    \item The LV and the SV  are on a main road (highway, rural or suburban road with dense traffic. Dense traffic is indicated, for example by  A traffic queue ahead that is visible to the SV driver, or can be inferred by the SV driver based on available information (e.g., traffic is moving significantly slower than the posted speed limit).
    \item Dense traffic in an adjacent lane with traffic in the same direction.
    \item The brake light had been turned on and off within a few seconds prior to the  unsurprising hard brake associated with the conflict  .
    \item The LV and the SV are on a road going downhill that requires application of the brakes.
    \item The LV is performing a narrow turn which requires slowing down.
    \item It can be anticipated that the LV is going to yield to another vehicle by slowing down (e.g., a vehicle in the parallel lane at a merge section).
\end{itemize}

\noindent
Conversely, a brake light onset is \textit{surprising} if it occurs in a situation where it is not unsurprising that theLV will slow down, such as:
\begin{itemize}
    \item The LV is in the middle of an controlled intersection after passing a green light.
    \item The SV is following the agent on a road with no blocking traffic visible to the SV driver.
\end{itemize}

2. \textit{Surprising agent deceleration}: Similar to brake light onset, agent deceleration is unsurprising if it occurs in a situation where the LV is expected to slow down (see examples under Note 1 above) and the magnitude of the deceleration is within the deceleration range typical for conflict-free (normal) driving. 
For present purposes, 3 m/$s^2$ can be used as a heuristic threshold for ``normal'' deceleration (for a passenger vehicle).

\noindent
Conversely, a lead vehicle deceleration is surprising if it is unusually hard (>3 m/$s^2$) and/or occurs in a situation where the LV is not expected to brake (see examples under Note 1 above).

3. \textit{Visibility of surprising agent deceleration}: LV deceleration is visible to the SV driver in terms of the looming (optical angular expansion rate) of the lead vehicle. At longer distances, the looming associated with the lead vehicle braking is considered not visible to the SV driver until the angular expansion rate exceeds 0.005 rad/s, which needs to be taken into account when determining first surprising agent deceleration. 

Visual looming is only directly indicative of LV braking if the SV and LV were initially driving roughly at the same speeds (i.e., constant headway). 
If the SV closes in on an LV moving at constant slower speed, looming (which is typically unsurprising by the SV driver) is produced and LV braking is indicated by an (surprising) change in the looming rate
If the LV is initially accelerating away from the SV (e.g., after overtaking) and then decelerates, looming will not occur until until the LV has reduced its speed to the SV's  speed

\begin{figure}[h]
	\centering
		\includegraphics[width=0.4\textwidth]{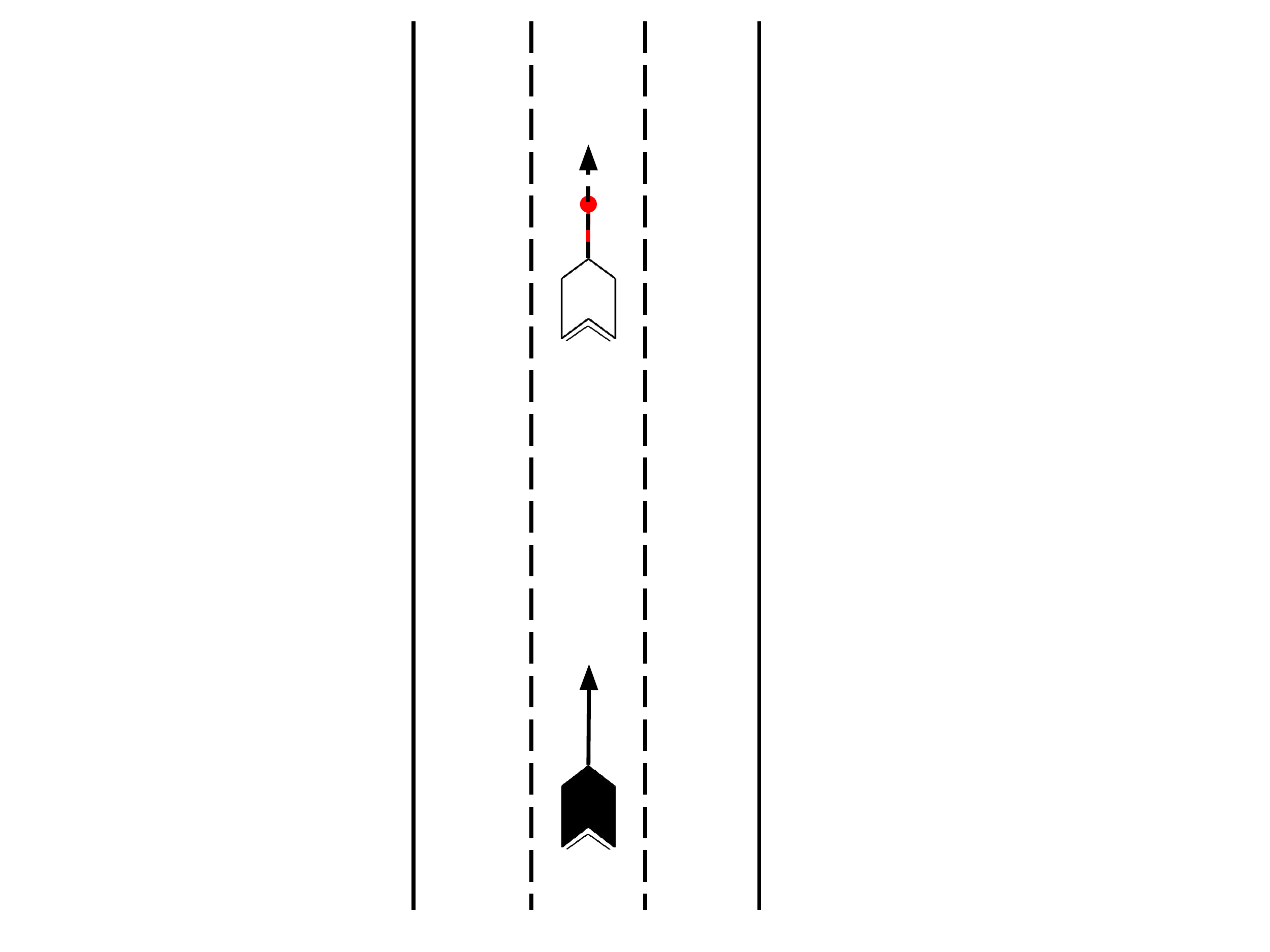}
	  \caption{ Pictogram illustrating scenario S.3.}
	  \label{appendix_fig3}
\end{figure}

\subsection*{S.2 The LV moves to exit the lane and then brakes surprisingly with the SV following behind}

The SV is driving at constant speed, the LV is making a 90-degree (or close to 90-degree) turn to exit the road, the LV brakes before finishing the turn and comes to stop in the SV's lane.

\noindent
Prior belief ($H_i$)
\begin{itemize}
    \item The LV  will complete the turn without stopping
\end{itemize}

\noindent
Surprise ($H_a$)
\begin{itemize}
    \item The LV brakes to a stop or near stop before completing the turn
\end{itemize}

\noindent
Stimulus onset ($T_1$)
\begin{itemize}
    \item Whatever occurs first: 
    \begin{itemize}
        \item First surprising LV deceleration that is visible to the SV  driver (e.g., the LV  starts braking harder than usual during the turn)
        \item First surprising brake light onset
        \item The LV comes to a complete stop
    \end{itemize}
\end{itemize}

\noindent
Stimulus end ($T_2$)
\begin{itemize}
    \item The first point in time at or after after $T_1$ where the angular expansion rate (looming) of the LV reaches 0.05 rad/s.
    \item  If the looming already reached 0.05 rad/s at $T_1$, or if it never reaches 0.05 rad/s (e.g. due to early braking or swerving), $T_2$ should be set to $T_1$.
\end{itemize}

\begin{figure}[t]
	\centering
		\includegraphics[width=0.2\textwidth]{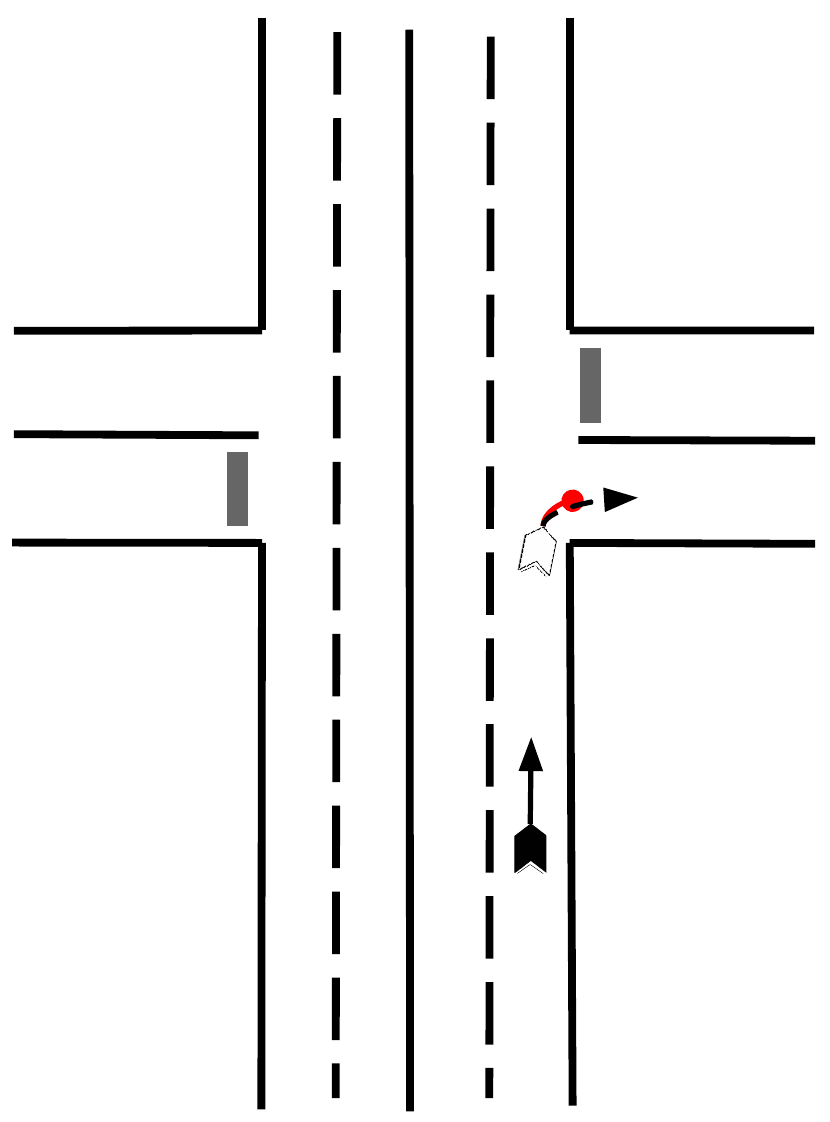}
	  \caption{ Pictogram illustrating scenario S.2.}
	  \label{appendix_s2}
\end{figure}

\subsection*{S.3 The LV  is stopped or slowing with the SV closing in from behind}

The SV approaches an LV that is stopped or driving at low speed ahead in a location where it would be surprising to stop (e.g., in the middle of a freeway, not close to an intersection or a traffic queue). Due to the long initial distance, and the absence of other cues that the agent is stopped, the SV driver believes that the LV is moving forward at higher speed and thus reacts late.

\noindent
Prior belief ($H_i$)
\begin{itemize}
    \item The LV  is moving forward
\end{itemize}

\noindent
Surprise ($H_a$)
\begin{itemize}
    \item The LV is stopped or slowing ahead
\end{itemize}

\noindent
Stimulus onset ($T_1$): 
\begin{itemize}
    \item First surprising looming visible to the SV  driver
\end{itemize}

\noindent
Stimulus end ($T_2$)
\begin{itemize}
    \item The first point in time at or after after T1 where the angular expansion rate (looming) of the LV reaches 0.05 rad/s.
    \item  If the looming already reached 0.05 rad/s at $T_1$, or if it never reaches 0.05 rad/s (e.g. due to early braking or swerving), $T_2$ should be set to $T_1$.
\end{itemize}

\noindent
NOTES
\begin{enumerate}
    \item In the absence of other cues, the closing between the LV and the SV can be detected by the SV driver in terms of the looming (optical angular expansion rate) of the LV.
Due to the small visual angle subtended by the LV at long distances, the closing is not visible to the SV driver until the angular expansion rate exceeds 0.005 rad/s.
\end{enumerate}

%% file: appendix_B.tex
\section{Generalization to other scenarios}\label{appendixC_generalizations}

\begin{figure}[h]
	\centering
		\includegraphics[width=0.4\textwidth]{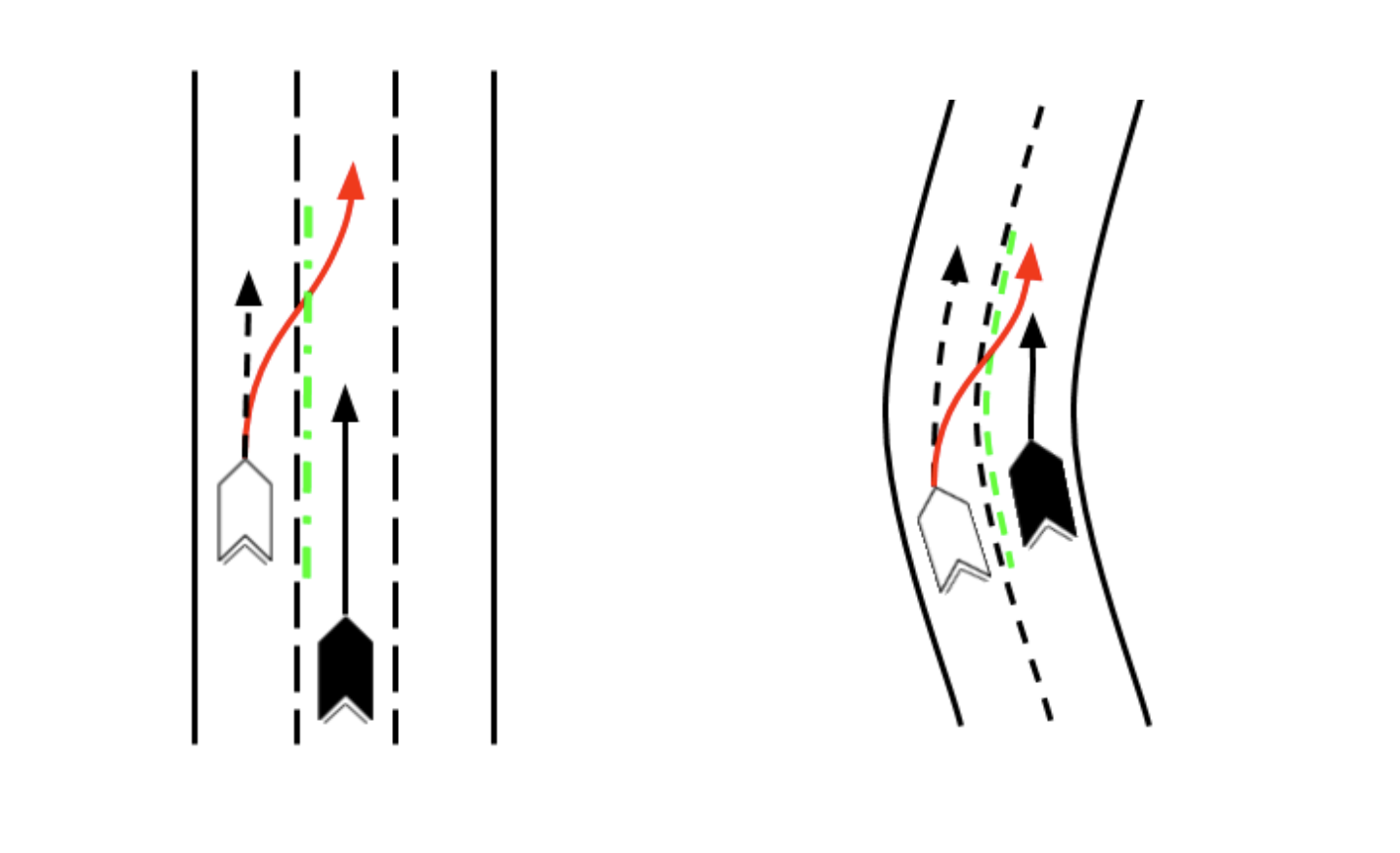}
	  \caption{Scenario for lateral cut-in from adjacent lane, same direction.}\label{appendix_c1}
\end{figure}

As the current implementation is based on manual annotation, it is straightforward to generalize the model to other use cases beyond lead vehicle braking scenarios. 
This is done by defining the initial (prior) and alternative (posterior) beliefs ($H_i$ and $H_a$) for each scenario of interest along with heuristics for annotating the stimulus onset ($T_1$) and stimulus end ($T_2$).
Some examples are given below:

\subsection*{Lateral cut-in from adjacent lane, same direction}
This scenario involves the principal other vehicle (POV)  cutting in front of the subject vehicle (SV) vehicle from an adjacent parallel lane in the same direction in a way that causes a conflict with the ego vehicle (Figure~\ref{appendix_c1}).
Here the prior belief is that the POV will continue straight in the original lane and the surprising posterior belief that the POV is cutting in front. 
If the vehicles are on a straight path, the stimulus onset $T_1$ (first surprising moment) can be defined as the moment when the POV starts moving laterally towards the SVs lane (Figure~\ref{appendix_c1}, left panel).
If the vehicles are on a curved path, $T_1$ is defined as the first moment when the agent's heading starts deviating (towards the ego vehicle) from the expected curved path defined by the curvature of the road (Figure~\ref{appendix_c1}, right panel). 
In both cases, the stimulus end  $T_2$ is defined by the moment the POV starts crossing the boundary of the ego vehicle's lane.

\subsection*{Crossing path scenario}

\begin{figure}[h]
	\centering
		\includegraphics[width=0.38\textwidth]{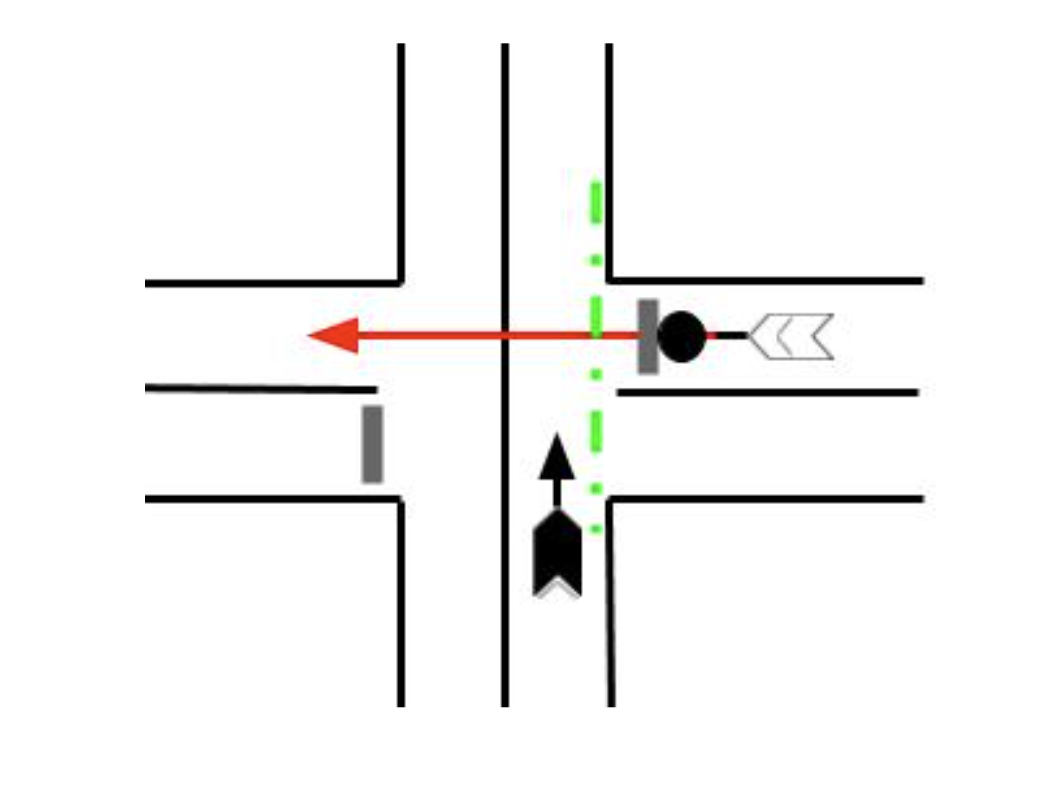}
	  \caption{Crossing path scenario.}\label{appendix_c2}
\end{figure}

This refers to a scenario where the SV has the right of way and the POV expected to yield (prior belief) but the POV surprisingly crosses into the SV's path thus causing a conflict (posterior belief; Figure~\ref{appendix_c2}).
Here the stimulus onset ($T_1$) is assigned as the first moment where the SV's behavior violates the prior belief, which can be operationalized by the following heuristics:

\begin{itemize}
    \item The POV is initially stationary or near stationary within 1 meter of the reference boundary and starts moving into the ego's lane with a speed of at least 1 m/s.
    \item The POV moves towards the intersection at constant speed and enters within 1 meter of the reference boundary without slowing down.
    \item The POV moves towards the intersection such that it would need to brake unusually hard (> 3 m/$s^2$) to stop short of the reference boundary.
    \item The POV appears from occlusion and has, at that moment, already satisfied one of the previous criteria.
\end{itemize}

The stimulus end ($T_2$) is assigned as the moment when the POV has crossed a reference boundary defined as the marked or virtual boundary of the SV's lane, or when the SV appears from occlusion and has, at this moment, already crossed the reference boundary.

Similar heuristics for determining $T_1$ and $T_2$, as the basis for computing ramp-up time, can in principle be established for any possible scenario, including other types of agents such as pedestrians and cyclists. 

\balance

%% file: appendix_C.tex
\section{Further details of visual looming computation}\label{appendixB}

To record the width of the lead vehicle in each frame of the video, we developed a computer vision solution for tracking the left and right corners of the vehicle and using the distance between the two as a measure of the optical vehicle width.
Tracking can be defined as locating an object across successive frames of a video, and in order to track the corners of the vehicle, a suitable algorithm was needed. 
To this end, OpenCV which is a library of programming functions suitable for real-time computer vision was used. Among OpenCV’s different object tracking implementations, CSRT (Discriminative Correlation Filter with Channel and Spatial Reliability, DCF-CSR) was used, because of the importance of accuracy for this particular problem.
Using the spatial reliability map, the filter support of the region of interest that is being tracked in each frame can be adjusted.
This improves tracking of the non-rectangular regions or objects and guarantees the enlarging and localization of the selected region. 
CSRT uses only 2 standard features (HoGs and Colornames) and operates at a comparatively lower fps (25 fps) while also giving higher accuracy for object tracking (for more information,